\documentclass[11pt]{article}
\usepackage[american]{babel}

\usepackage[round]{natbib}
\usepackage{xifthen}
\usepackage{mathtools}
\usepackage{booktabs}
\usepackage{tikz}
\usepackage{array}
\usepackage[a4paper]{geometry}
\usepackage{multirow}
\usepackage{amsmath, amsthm}
\usepackage{subcaption}
\usepackage{bm}
\newtheorem{thm}{Theorem}
\newtheorem{lem}{Lemma}

\newcolumntype{C}[1]{>{\centering\arraybackslash}m{#1}}
\newtheorem{defin}{Definition}
\DeclareMathOperator{\val}{val}
\DeclareMathOperator{\rank}{rank}
\DeclareMathOperator{\paOP}{pa}
\DeclareMathOperator{\OR}{OR}
\newcommand{\pa}[1][]{%
\ifthenelse{ \equal{#1}{} }
{\paOP}
{\paOP_{#1}}
}
\newcommand{\+}[1]{\ensuremath{\mathbf{#1}}}
\newcommand{\eqs}{\!=\!}
\newcommand{\sG}{\mathcal{G}}
\usetikzlibrary{positioning, arrows.meta, shapes.geometric, shapes.misc}
\tikzset{%
  semithick,
  >={Stealth[width=1.5mm,length=2mm]},
  obs/.style 2 args = {name = #1, circle, inner sep = 8pt, label = center:$#2$}
}
\newcommand\independent{\protect\mathpalette{\protect\independenT}{\perp}}
\def\independenT#1#2{\mathrel{\rlap{$#1#2$}\mkern2mu{#1#2}}}
\newcommand{\cond}{\,\vert\,}

\title{Full Law Identification under Missing Data with Categorical Variables}

\author{Santtu Tikka and Juha Karvanen}
\date{%
  Department of Mathematics and Statistics\\
  University of Jyvaskyla\\
  Finland
}

\begin{document}
\maketitle

\begin{abstract}
Missing data may be disastrous for the identifiability of causal and statistical estimands. In graphical missing data models, colluders are dependence structures that have a special importance for identification considerations. It has been shown that the presence of a colluder makes the full law, i.e., the joint distribution of variables and response indicators, non-parametrically non-identifiable. However, when the variables related to the colluder structure are categorical, it is sometimes possible to regain the identifiability of the full law. We present a necessary and sufficient condition for the identification of the full law in the presence of colluder structures with arbitrary categorical variables. Maximum likelihood estimation of the full law in identifiable models with categorical variables is demonstrated with simulated and real data.

\end{abstract}

\section{Introduction}\label{sec:intro}
Missing data poses a serious threat to the validity of inference because it leads to information loss and may cause an estimation task to become infeasible due to non-identifiable estimands. Missing observations are almost unavoidable in survey data and are commonly encountered in other data sources as well. Traditionally, missing data problems have been approached with the classification missing completely at random (MCAR), missing at random (MAR), and missing not at random \citep[MNAR,][]{rubin1976inference,seaman2013meant,little2019statistical}.

Graphical missing data models \citep{fay1986causal,NIPS2013_0ff8033c,Karvanen2015studydesign} are not tied to the MCAR-MAR-MNAR classification but describe the missing data mechanism variable by variable. A missing data graph consists of fully observed variables, partially observed variables and their response indicators, and observed proxy variables that have some missing values. The edges of the graph indicate causal relations between these variables in a non-parametric form. A distribution compatible with the graph is said to be faithful if all conditional independence properties of the distribution are implied by the graph.

With graphical missing data models, the identification of causal or statistical estimands may be considered separately from the estimation. One of the main questions of interest has been the identification of the \emph{full law}, i.e., the joint distribution of fully observed variables, and partially observed variables and their response indicators, from the available distribution of the observed data. \citet{NIPS2013_0ff8033c} referred to this question as \emph{recoverability} and provided several specific conditions for identifiability.

Various algorithms for identifiability under missing data have been proposed, which turned out to not provide a complete characterization of the problem \citep{NIPS2014_31839b03, 10.5555/3020847.3020930, pmlr-v77-tian17a, pmlr-v115-bhattacharya20b, JSSv099i05}. Ultimately, a sound and complete identifiability criterion for the full law was provided by \citet{pmlr-v119-nabi20a}.

Curiously, when missing data is involved, even the identifiability of purely observational quantities becomes a causal inference problem. This can be demonstrated by a simple example where the interest is identifying the full law, but the observed distribution has no conditional independence properties that could be exploited to obtain the identifying functional, and instead, the target quantity is obtained via do-calculus; the primary tool of inference for identifying interventional distributions in structural causal models \citep{Mohan_2021,pearl_2009}.

The identifiability criterion for the full law in a general non-parametric setting is tied to the presence of self-censoring, which in this context means that a variable is a direct cause of its response indicator, and to the existence of dependence structures called \emph{colluders} \citep{pmlr-v115-bhattacharya20b,pmlr-v119-nabi20a}. In a colluder structure, which was considered already by \citet[Figure~\ref{fig:colluder}]{fay1986causal} but without a specific name, the missingness of a variable is caused by another variable and its response indicator. The no self-censoring model \citep{sadinle2017itemwise, shpitser2016consistent, malinsky2022semiparametric} is a notable identifiable special case where the data are MNAR but self-censoring and colluders are not present.

While colluder structures have been considered in missing data literature, their  importance may have gone largely unnoticed as their role in full law identification was characterized only quite recently by \citet{pmlr-v115-bhattacharya20b}. Colluder structures are not only interesting theoretically but are also  relevant for practical applications. For instance, in clinical practice, the decision to conduct a diagnostic test may depend both on the availability of an earlier test result and the actual result of this earlier test. In surveys, the order of questions is important and the participant's choice to answer or not a certain question may depend on the previous questions and whether these questions were answered.

\citet{li2023self} considered MNAR scenarios with self-censoring but without colluders and showed that a completeness \citep{lehmann1950completeness} assumption is sufficient to regain the identifiability of the full law. We focus on complementary scenarios without self-censoring but with colluders. Besides these scenarios, %self-censoring scenario of \citep{li2023self}, 
there are also other cases where additional assumptions can help to identify the full law when it is non-identifiable in general. Sufficient information for identifiability can be obtained e.g., from other variables that are fully observed \citep{miao2023identification,ma2003identification}, from assumptions on the variable types \citep{ma2003identification}, or from parametric or semi-parametric assumptions \citep{zhou2010block,tchetgen2018discrete}.

Graphical missing data models where the variables are categorical form an important special case with practical relevance. For instance, in surveys, the variables are typically categorical. The number of parameters needed to fully specify these models depends on the structural restrictions implied by the graph and the number of categories of each variable. Incomplete categorical data have often been analyzed with log-linear models \citep{fay1986causal,clarke2002boundary,kim2020log} where additional restrictions can be set e.g., by excluding the interactions between some variables. We do not make any additional assumptions besides assuming that the variables involved with the colluder structures are categorical and a standard positivity assumption such that all conditional distributions are well-defined. The variables outside the colluder structures can be continous or discrete.

In this paper, we consider the identification of the full law in graphical missing data models in the presence of colluder structures with categorical variables. We first show that an earlier construction used to prove non-identifiability when colluders are present \citep{pmlr-v115-bhattacharya20b,pmlr-v119-nabi20a} is not applicable if the variables in the colluder structure are dependent. We then present an alternative construction for this setting under the assumption of dependence. Finally, as the main contribution, we present a necessary and sufficient condition for the identification of the full law when colluder structures are present with arbitrary categorical variables. The condition is related to a system of linear equations and can be expressed as a full-rank requirement for a matrix. The presented approach is similar to identification strategies with proxy variables \citep[see e.g.,][]{Kuroki_2014, Miao_2018, phung2024zero}.

The rest of the paper is organized as follows. The notation is introduced in Section~\ref{sec:notation}. In Section~\ref{sec:models}, the existing non-identifiability results regarding the colluder structure are revisited with an additional assumption of dependency. In Section~\ref{sec:ccm}, categorical colluder models are formally defined and the necessary and sufficient condition for full law identifiability in these models is presented. Section~\ref{sec:examples} considers example scenarios including unmeasured confounding. Estimation of the full law for categorical colluder models is demonstrated in Section~\ref{sec:simulation} with simulated data, and with real data in Section~\ref{sec:application}. Section~\ref{sec:conclusion} concludes the paper.

\section{Notation and Definitions}\label{sec:notation}

We will use the following notation. Capital letters denote random variables or vertices, and small letters denote values of random variables. Bold letters are used to denote sets and vectors of random variables, vertices, or values. We assume that the reader is familiar with basic graph theoretical concepts such as vertices, edges, and paths.

A directed acyclic graph (DAG) $\sG$ is a pair $(\+ V, \+ E)$ where $\+ V$ is a set of vertices and $\+ E$ is a set of directed edges such that there are no cycles, i.e., directed paths from $X$ to itself, in the graph. DAGs are commonly used to depict relationships between random variables of interest, for example in Bayesian networks and structural causal models \citep{pearl_2009}. A joint distribution compatible with a DAG $\sG$ satisfies the following factorization
\begin{equation} \label{eq:factorization}
  p(\+ V) = \prod_{V_i \in \+ V} p(V_i \cond \pa[\sG](V_i)),
\end{equation}
where $\pa[\sG](V_i)$ denotes the parents of $V_i$ in $\sG$. This factorization enables conditional independence statements about members of $\+ V$ to be derived from the DAG based on the well-known d-separation criterion \citep{981e909a-69b6-346e-bd9d-1ef8e392bda3,pearl_2009}. Additionally, if the distribution is assumed to be \emph{faithful} to the DAG $\sG$, then the statements implied by d-separation are exactly those that are implied by $p(\+ V)$. We will denote d-separation statements and the implied conditional independence relations such as $\+X$ is independent of $\+Y$ given $\+Z$ as $\+X \independent \+Y \cond \+Z$.

In many practical scenarios, in addition to variables that contain missing data, there may exist variables that are completely unobserved, but that are nonetheless relevant to the research question at hand. We can encode the dependency structure of the observed variables $\+ V$ and the unobserved variables $\+ U$ using a hidden variable DAG over $\+ V \cup \+ U$. However, as we only have access to $p(\+ V)$, there are infinitely many hidden variable DAGs that encode the same set of conditional independence relations between variables in $\+ V$. The typical approach is to instead consider acyclic directed mixed graphs \citep[ADMG,][]{richardson_2003, 10.1214/22-AOS2253} where arbitrary structures of the unobserved variables are replaced by bidirected edges, corresponding to unmeasured confounding between pairs of observed variables. Importantly, no information on the dependence structure is lost if a hidden variable DAG is converted to a corresponding ADMG as the resulting ADMG entails the same conditional independence properties between the observed variables as the original hidden variable DAG \citep{10.1214/17-AOS1631}. This ADMG is also unique and can be obtained via the latent projection \citep{verma1990}. In ADMGs, the d-separation condition is replaced by the analogous m-separation condition that accounts for the bidirected edges \citep{richardson_2003}.

\section{Missing Data Models} \label{sec:models}

In missing data models, the random variables $\+ V$ can be partitioned as $\+ V = \{\+ O, \+ X^{(1)}, \+ X, \+ R\}$, following the convention by \citet{NIPS2013_0ff8033c}. The set of completely observed variables is denoted by $\+ O$, the set of partially observed variables is denoted by $\+ X^{(1)}$, the set of binary response indicators of the variables in $\+ X^{(1)}$ (also called missingness indicators) is denoted by $\+ R$, and the set of observed proxy variables of $\+ X^{(1)}$ is denoted by $\+ X$. We will also refer to members of $\+ X^{(1)}$ as true variables, for short. Each true variable has a corresponding response indicator and a corresponding observed proxy variable. The response indicators, observed proxies, and the true variables have the following functional relationship by definition:
\begin{equation} \label{eq:missingness_mechanism}
  X_i = \begin{cases}
    X_i^{(1)} & \text{if } R_i = 1, \\
    \text{NA} & \text{if } R_i = 0,
  \end{cases}
\end{equation}
where \text{NA} (not available) denotes a missing value. In other words, the true variable is observed only if it is not missing as indicated by the response indicator. A missing data model is a set of distributions over $\+ V$ that satisfies the above definition. Because the observed proxies are determined deterministically from the true variables and the response indicators, the missing data distribution reduces to the \emph{full law} $p(\+ O, \+ X^{(1)}, \+ R)$. The full law consists of two parts: the \emph{target law} $p(\+ O, \+ X^{(1)})$ and the \emph{missingness mechanism} $p(\+R \cond \+ O, \+ X^{(1)})$. Finally, the \emph{observed data distribution} $p(\+ O, \+ X, \+ R)$ is the incomplete version of the full law, which represents the available information under missing data.

We call DAGs with compatible missing data models \emph{missing data DAGs} (and analogously \emph{missing data ADMGs} when hidden variables are present). Such DAGs must satisfy the following additional restrictions due to equation~\eqref{eq:missingness_mechanism}: Members of $\+ R$ cannot be parents of members of $\+ O \cup \+ X^{(1)}$, the set of parents of each $X_i \in \+ X$ is $\{X^{(1)}_i, R_i\}$, and members of $\+ X$ do not have children. We will denote missing data DAGs with vertex set $\+ V$ by $\sG(\+ V)$ or $\sG(\+ O, \+ X^{(1)}, \+ X, \+ R)$.

Non-parametric identification in missing data models has two goals: the identification of full law $p(\+ O, \+ X^{(1)}, \+ R)$ and the identification of the target law $p(\+ O, \+ X^{(1)})$ in terms of the observed data law $p(\+ O, \+ X, \+ R)$ (or some functionals of them). The identifiability of the full law is equivalent to the identifiability of the missingness mechanism because we can write
\begin{equation} \label{eq:full_law_chainrule}
  p(\+ O, \+ X^{(1)}, \+ R) = \frac{p(\+ O, \+ X^{(1)}, \+ R = 1)}{p(\+ R = 1 \cond \+ O, \+ X^{(1)})}p(\+ R \cond \+ O, \+ X^{(1)}),
\end{equation}
where $\+ R = 1$ means that all response indicators have a value assignment of 1, and $p(\+ O, \+ X^{(1)}, \+ R = 1)$ is directly identifiable from the observed data distribution. In this paper, we will focus on full law identification via identification of the missingness mechanism as licensed by equation~\eqref{eq:full_law_chainrule}. \citet{pmlr-v119-nabi20a} provided a necessary and sufficient graphical criterion for the identification of the full law in missing data DAGs. This criterion involves an important structure called a colluder.

\begin{figure}[t]
  \begin{center}
    \begin{subfigure}{0.25\textwidth}
      \begin{center}
        \begin{tikzpicture}[scale=1.5]
        \node [obs = {X}{X^{(1)}}] at (0,1) {};
        \node [obs = {Y}{Y^{(1)}}] at (1,1) {};
        \node [obs = {RX}{R_X}] at (0,0) {};
        \node [obs = {RY}{R_Y}] at (1,0) {};
        \node [obs = {X*}{X}] at (0,-1) {};
        \node [obs = {Y*}{Y}] at (1,-1) {};
        \draw[->, lightgray] (X) to[bend right=25] (X*);
        \draw[->, lightgray] (Y) to[bend left=25] (Y*);
        \draw[->] (X) -- (RY);
        \draw[->] (RX) -- (RY);
        \draw[->, lightgray] (RX) -- (X*);
        \draw[->, lightgray] (RY) -- (Y*);
        \end{tikzpicture}
      \end{center}
      \caption{}
      \label{fig:colluder1}
    \end{subfigure}
    \hspace{0.25cm}
    \begin{subfigure}{0.25\textwidth}
      \begin{center}
        \begin{tikzpicture}[scale=1.5]
        \node [obs = {X}{X^{(1)}}] at (0,1) {};
        \node [obs = {Y}{Y^{(1)}}] at (1,1) {};
        \node [obs = {RX}{R_X}] at (0,0) {};
        \node [obs = {RY}{R_Y}] at (1,0) {};
        \node [obs = {X*}{X}] at (0,-1) {};
        \node [obs = {Y*}{Y}] at (1,-1) {};
        \draw[->] (X) -- (Y);
        \draw[->,lightgray] (X) to[bend right=25] (X*);
        \draw[->,lightgray] (Y) to[bend left=25] (Y*);
        \draw[->] (X) -- (RY);
        \draw[->] (RX) -- (RY);
        \draw[->, lightgray] (RX) -- (X*);
        \draw[->, lightgray] (RY) -- (Y*);
        \end{tikzpicture}
      \end{center}
      \caption{}
      \label{fig:colluder2}
    \end{subfigure}
  \end{center}
  \caption{Missing data DAGs that contain a colluder structure $X^{(1)} \rightarrow R_Y \leftarrow R_X$. Edges that correspond to deterministic relationships are shown in gray. For binary $X^{(1)}$ and binary $Y^{(1)}$, the full law is identifiable in (b) if $Y^{(1)}$ depends on $X^{(1)}$, but not identifiable in (a).}
  \label{fig:colluder}
\end{figure}
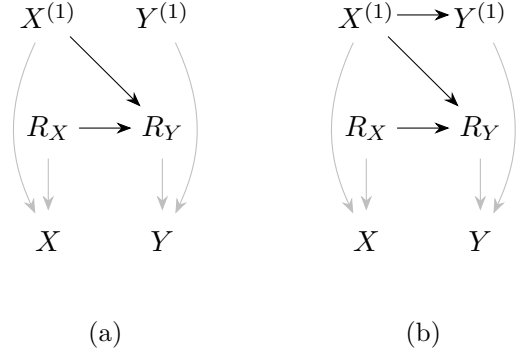

\begin{defin}[Colluder] A set $\{X^{(1)}, R_X\}$ is a colluder of $R_Y$ in a missing data DAG $\sG(\+ O, \+ X^{(1)}, \+ R, \+ X)$ if $\sG$ contains the edges $X^{(1)} \rightarrow R_Y$ and $R_X \rightarrow R_Y$.
\end{defin}

For a colluder $\{X^{(1)}, R_X\}$ of $R_Y$, we say that $X^{(1)}$ and $R_X$ are \emph{colluding}. We will also depict colluders in graphical terms, for example, Figure~\ref{fig:colluder1} has a colluder $X^{(1)} \rightarrow R_Y \leftarrow R_X$. One of the conditions in the graphical criterion of \citet{pmlr-v119-nabi20a} states that no colluders should be present in the missing data DAG. A colluder prevents the identifiability of the corresponding term $p(R_Y|\pa[\sG](R_Y))\rvert_{R_X = 0}$ in the factorization of equation~\eqref{eq:factorization}, thus preventing the identifiability of the missingness mechanism $p(\+R \cond \+ O, \+ X^{(1)})$ and %consequently 
the full law $p(\+ O, \+ X^{(1)}, \+ R)$. The presence of colluder structures was shown to render the full law non-identifiable by the following lemma of \citet{pmlr-v115-bhattacharya20b}.

\begin{lem}[\citet{pmlr-v115-bhattacharya20b}, Lemma~1] \label{lem:colluder_lemma}
In a missing data DAG $\sG(\+ O, \+ X^{(1)}, \+ R, \+ X)$, if there exists $R_i, R_j \in \+ R$ such that $\{R_j, X^{(1)}_j\} \in \pa[\sG](R_i)$, then $p(R_i|\pa[\sG](R_i))\rvert_{R_j = 0}$ is not identified. Hence, the full law $p(\+ O, \+ X^{(1)}, \+ R)$ is not identified.
\end{lem}

However, the construction provided as the proof of Lemma~\ref{lem:colluder_lemma} by \citet{pmlr-v115-bhattacharya20b} is no longer applicable if a single edge is added to the missing data DAG used in the construction and if we assume that this edge corresponds to a true dependency in the full law. The proof of the lemma considers the missing data DAG shown in Figure~\ref{fig:colluder1} and the provided construction is also applicable in the missing data DAG of Figure~\ref{fig:colluder2} where the parametrization provided simply assumes that the edge $X^{(1)} \rightarrow Y^{(1)}$ has no effect, i.e., the conditional distribution $p(Y^{(1)} \cond X^{(1)})$ is the same regardless of the value of $X^{(1)}$. This example was also considered by \citet{hunt_phd}.

\subsection{Identifiable Binary Colluder Model} \label{sec:example1}

As our first example of identification with colluders, we consider the missing data DAG of Figure~\ref{fig:colluder2} where the edge $X^{(1)} \rightarrow Y^{(1)}$ is present and we assume that $p(Y^{(1)} \cond X^{(1)})$ depends on the value of $X^{(1)}$. We note that this assumption is much weaker than faithfulness because it considers only one specific conditional distribution. In order to describe the dependency of $X^{(1)}$ and $Y^{(1)}$, we must include an additional parameter in the construction. As a consequence, the full law becomes identifiable via the following identifying functional for $P(R_Y = 1 \cond X^{(1)}, R_X = 0)$:
\begin{equation} \label{eq:binarysolution}
  \begin{aligned}
    p(R_Y = 1 \cond X^{(1)} = 0, R_X = 0) &= \frac{r - h r - h s}{ab(c - h)}, \\
    p(R_Y = 1 \cond X^{(1)} = 1, R_X = 0) &= \frac{r - c r - c s}{a(b-1)(c-h)},
  \end{aligned}
\end{equation}
where
\begin{equation} \label{eq:binaryvariableroles}
  \begin{aligned}
    a &= p(R_X \eqs 0), \\
    b &= p(X^{(1)} \eqs 0), \\
    c &= p(Y^{(1)} \eqs 0 \cond X^{(1)} \eqs 0), \\
    h &= p(Y^{(1)} \eqs 0 \cond X^{(1)} \eqs 1), \\
    r &= p(Y \eqs 0, R_X \eqs 0, R_Y \eqs 1), \\
    s &= p(Y \eqs 1, R_X \eqs 0, R_Y \eqs 1),
  \end{aligned}
\end{equation}
and the assumption ${p(Y^{(1)} = 0 \cond X^{(1)} = 0) \neq p(Y^{(1)} = 0 \cond X^{(1)} = 1)}$ guarantees that the denominators in Equation~\eqref{eq:binarysolution} are not zero ($c \neq h$) assuming positivity for $a,b,c,$ and $h$. The solution~(\ref{eq:binarysolution}) can be written in a matrix form
\begin{equation} \label{eq:binarymatrix}
  \begin{pmatrix}
    p(R_Y = 1 \cond X^{(1)} = 0, R_X = 0) \\
    p(R_Y = 1 \cond X^{(1)} = 1, R_X = 0)
  \end{pmatrix}
  = \+A^{-1} \+b,
\end{equation}
where
\[
  \begin{aligned}
    & \+A = 
    \begin{pmatrix}
      abc & a(1-b)h \\
      ab(1-c) & a(1-b)(1-h)
    \end{pmatrix}, \quad
    \+b = 
    \begin{pmatrix}
      r \\
      s
    \end{pmatrix}, \\
    & \+A^{-1} = \frac{1}{c-h}\begin{pmatrix}
      \dfrac{1-h}{ab}      & -\dfrac{h}{ab} \\[9pt]
      -\dfrac{1-c}{a(1-b)} & \dfrac{c}{a(1-b)}
    \end{pmatrix}. 
  \end{aligned}
\]
The quantities $a,b,c,h,r,s$ in equation~(\ref{eq:binaryvariableroles}) are directly identifiable from the observed data distribution $p(X,Y,R_X,R_Y)$ by noting that $X^{(1)}$ is independent of its response indicator $R_X$ and similarly $Y^{(1)}$ is independent of the response indicators $R_Y$ and $R_X$ given $X^{(1)}$. Thus the full law is identifiable. The complete details of the construction and derivation of the identifying functional can be found in the Supplementary Material. The parameter names used in the construction are chosen to match those used by \citet{pmlr-v115-bhattacharya20b} with the new additional parameter $h$ related to the edge $X^{(1)} \rightarrow Y^{(1)}$. Intuitively, we are able to identify the full law because the connection between $X^{(1)}$ and $Y^{(1)}$ allows $Y^{(1)}$ to reveal information about $X^{(1)}$ that would otherwise not be possible to obtain due to the missingness of $X^{(1)}$. We note that while we assumed a dependency between $X^{(1)}$ and $Y^{(1)}$ via the edge $X^{(1)} \rightarrow Y^{(1)}$ in this example, an analogous proof of identifiability can be derived under any structure of dependency between these two variables, such as $Y^{(1)} \rightarrow X^{(1)}$ or a hidden common cause $X^{(1)} \leftarrow U \rightarrow Y^{(1)}$, as long as the dependency is assumed to hold in the full law.

\subsection{Non-identifiable Ternary Colluder Model} \label{sec:example2}

As our second example, we consider again the missing data DAG of Figure~\ref{fig:colluder2} but assume instead that $X^{(1)}$ is ternary while $Y^{(1)}$ is binary. Now it is once again possible to provide a construction that shows non-identifiability, even under the dependency assumption. The main idea of the construction is that the conditional distribution of $Y^{(1)}$ is defined to be exactly the same for two distinct values of $X^{(1)}$. This makes it impossible to learn about the distribution of $X^{(1)}$ by observing $Y$. The details of the construction are given in the Supplementary Material. This construction proves Lemma~\ref{lem:colluder_lemma} in the situation of Figure~\ref{fig:colluder2} under the assumption on the dependency of $X^{(1)}$ and $Y^{(1)}$.

\section{Colluders and Categorical Variables} \label{sec:ccm}

The examples of the previous section suggest that the number of categories of the true variables that correspond to the two response indicators present in the colluder structure is a crucial factor of identifiability. Motivated by these examples, we will show that not only do colluders sometimes permit identifiability of the full law with binary variables, but they also do so for a large class of categorical models. For this purpose, we define the \emph{categorical colluder model}.

\begin{defin}[Categorical colluder model]
A missing data model is a categorical colluder model $\textrm{CCM}(m,q)$ if the missing data DAG contains a colluder $\{X^{(1)}, R_X\}$ of $R_Y$, and $X^{(1)}$ and $Y^{(1)}$ are categorical variables with $m$ and $q$ classes, respectively.
\end{defin}

Moreover, we say that $\textrm{CCM}(m,q)$ is a categorical colluder model with respect to the colluder $\{X^{(1)}, R_X\}$ of $R_Y$. Naturally, this definition allows a single missing data model to be a categorical colluder model with respect to several colluders simultaneously, with possibly different numbers of categories in each colluder structure. Variables not related to the colluder structures can be discrete or continuous.

\subsection{Colluder Equations}

In a categorical colluder model, we can construct special equations that play a key role in full law identification.

\begin{lem}[Colluder equations] \label{lem:colluder_equations}
Let $\textrm{CCM}(m,q)$ be a categorical colluder model with respect to a colluder $\{X^{(1)}, R_X\}$ of $R_Y$. Then
\[
\begin{aligned}
  &p(Y^{(1)} \eqs y_k, R_Y \eqs 1, R_X \eqs r \cond \+ Z)= \\
  &\sum_{j = 1}^m p(Y^{(1)} \eqs y_k \cond R_X \eqs r, X^{(1)} \eqs x_j, R_Y \eqs 1, \+ Z)p(R_X \eqs r, X^{(1)} \eqs x_j \cond R_Y \eqs 1, \+ Z)p(R_Y \eqs 1 \cond \+ Z)
\end{aligned}
\]
where $\+Z = (\+ O, \+ X^{(1)} \setminus \{Y^{(1)}, X^{(1)} \}, \+ R \setminus \{R_X, R_Y\})$. The above equation is called the colluder equation of the class $Y=y_k$ for $R_X = r$. Colluder equations for all classes of $Y^{(1)}$ and $R_X = r$ are expressed collectively in matrix form as
\[
\+A_r \+s_r = \+b_r
\]
where
\[
\+s_r =
\begin{pmatrix}
p(R_X \eqs r, X^{(1)} \eqs x_1 \cond R_Y \eqs 1, \+ Z) \\
\vdots \\
p(R_X \eqs r, X^{(1)} \eqs x_m \cond R_Y \eqs 1, \+ Z)
\end{pmatrix},
\]
is the vector of probabilities to be solved,
\[
\+A_r =
\begin{pmatrix}
a_{r11} & \ldots & a_{r1m} \\
 \vdots & \ddots & \vdots \\
a_{rq1} & \ldots & a_{rqm}
\end{pmatrix},
\]
is the colluder matrix such that
\[
  a_{rij} = p(Y^{(1)} \eqs y_i \cond R_X \eqs r, X^{(1)} \eqs x_j, R_Y \eqs 1, \+ Z)p(R_Y \eqs 1 \cond \+ Z)
\]
where $i = 1,\ldots,q$, $j = 1,\ldots,m$ and
\[
\+b_r =
\begin{pmatrix}
p(Y^{(1)} \eqs y_1, R_Y \eqs 1, R_X \eqs r \cond \+ Z) \\
\vdots \\
p(Y^{(1)} \eqs y_q, R_Y \eqs 1, R_X \eqs r \cond \+ Z)
\end{pmatrix}
\]
is the right-hand side of the colluder equations.
\end{lem}
Proof. The equations follow by applying the law of total probability to the left-hand side term ${p(Y^{(1)} \eqs y_k, R_Y \eqs 1, R_X \eqs r \cond \+ Z)}$ with respect to $X^{(1)}$ and then factorizing the joint distribution. In order to take advantage of the colluder equations, we first prove a utility lemma.

\begin{lem} \label{lem:colluder_eqs_id}
Let $\textrm{CCM}(m,q)$ be a categorical colluder model with respect to a colluder $\{X^{(1)}, R_X\}$ of $R_Y$ and let ${\+A_r \+s_r = \+b_r}$ be the corresponding colluder equations for $r \in \{0,1\}$. If ${R_X \independent Y^{(1)} \cond \+ O, \+ X^{(1)} \setminus Y^{(1)}, \+ R \setminus R_X}$ and elements of $\+ R \cap \+ Z$ have the value assignment of 1, then all elements of the colluder matrix $\+ A_r$ are identifiable and all elements of the right-hand side vector $\+ b_r$ are identifiable.
\end{lem}
Proof. Let us first consider the case where $r = 0$. The left-hand side 
\[ 
  p(Y^{(1)} \eqs y_k, R_Y \eqs 1, R_X \eqs 0 \cond \+ Z)\rvert_{(\+ R \cap \+ Z) = 1}
\] 
is directly identifiable from the observed data distribution. Using the assumed conditional independence of $R_X$ and $Y^{(1)}$ given the other variables, we can write the first term on the right-hand side as %identify the first term on the right-hand side by writing
\[
  p(Y^{(1)} \eqs y_i \cond R_X \eqs 0, X^{(1)} \eqs x_j, R_Y \eqs 1, \+Z) = p(Y^{(1)} \eqs y_i \cond R_X \eqs 1, X^{(1)} \eqs x_j, R_Y \eqs 1, \+ Z),
\]
and we can identify $p(Y^{(1)} \eqs y_i \cond R_X \eqs 1, X^{(1)} \eqs x_j, R_Y \eqs 1, \+ Z)\rvert_{(\+ R \cap \+ Z) = 1}$ from the observed data distribution. The second term $p(R_Y \eqs 1 \cond \+ Z)\rvert_{(\+ R \cap \+ Z) = 1}$ is directly identifiable regardless of the value of $R_X$. In the case where $r = 1$, all terms of $\+ A$ and $\+ b$ are directly identifiable from the observed data distribution. 

Next, we show how the identifiability of the colluder equations is connected to the identifiability of the missingness mechanism.

\subsection{Identification of the Missingness Mechanism}

Consider a colluder $\{X^{(1)}, R_X\}$ of $R_Y$. Recalling Lemma~\ref{lem:colluder_lemma}, a straightforward strategy for the identification of the missingness mechanism would be to attempt to regain identifiability of the term rendered non-identifiable by the colluder, i.e., the conditional distribution $p(R_Y|\pa[\sG](R_Y))\rvert_{R_X = 0}$, as we did in our first example in Section~\ref{sec:example1}. While it may seem intuitive to attempt to identify this ``missing piece'' in the presence of colluders, and even though it is possible to derive similar equations as the colluder equations of Lemma~\ref{lem:colluder_equations} to identify $p(R_Y|\pa[\sG](R_Y))\rvert_{R_X = 0}$, this approach does not provide a general method for full law identification.

In contrast, the conditional distribution $p(R_X \cond \+ O, \+ X^{(1)}, (\+ R \setminus R_X) = 1)$ is directly linked to full law identification via the odds ratio (OR) parameterization of the missingness mechanism: 
\begin{equation} \label{eq:or_factorization}
  p(\+ R \cond \+ O, \+ X^{(1)}) = \frac{1}{Z} \prod_{k=1}^K p(R_k \cond \+ R_{-k} \eqs 1, \+ O, \+ X^{(1)}) \prod_{k=2}^K \OR(R_k, \+ R_{\prec k} \cond \+ R_{\succ k} \eqs 1, \+ O, \+ X^{(1)}),
\end{equation}
where $\+ R_{-k} = \+ R \setminus R_k$, $\+ R_{\prec k} = \{R_1, \ldots, R_{k-1}\}$, $\+ R_{\succ k} = \{R_{k+1}, \ldots R_K\}$, and
\[
  \begin{aligned}
  &\OR(R_k, \+ R_{\prec k} \cond \+ R_{\succ k} \eqs 1, \+ O, \+ X^{(1)}) \\
  &=\frac{p(R_k \cond \+ R_{\succ k} \eqs 1, \+ R_{\prec k}, \+ O, \+ X^{(1)})}{p(R_k \eqs 1 \cond \+ R_{\succ k} \eqs 1, \+ R_{\prec k}, \+ O, \+ X^{(1)})}
  \frac{p(R_k \eqs 1 \cond \+ R_{-k} \eqs 1, \+ O, \+ X^{(1)})}{p(R_k \cond \+ R_{-k} \eqs 1, \+ O, \+ X^{(1)})},
  \end{aligned}
\]
and $Z$ is the normalizing constant. \citet{pmlr-v119-nabi20a} showed that if no colluders or self-censoring edges (i.e., edges of the form $X^{(1)} \rightarrow R_X$) are present, then the terms $p(R_k \cond \+ R_{-k} \eqs 1, \+ O, \+ X^{(1)})$ in the above factorization are all identifiable due to the conditional independence of $R_k$ and $X_k^{(1)}$ given all other fully and partially observed variables and all other response indicators (the OR terms are always identifiable regardless of colluders or self-censoring edges). If a colluder $\{X^{(1)}, R_X\}$ of $R_Y$ is present, the corresponding conditional independence no longer holds as $R_X$ is a member of the Markov-blanket of $X^{(1)}$. Thus we must find an alternative strategy to identify the corresponding conditional distribution $p(R_X \cond \+ O, \+ X^{(1)}, (\+ R \setminus R_X) = 1)$. We do this via the same conditional independence relation that we used to solve the colluder equations.

\begin{thm} \label{thm:CCMidentifiability}
Let $\sG$ be the missing data DAG of a categorical colluder model $\textrm{CCM}(m,q)$ with respect to a colluder $\{X^{(1)}, R_X\}$ of $R_Y$ and {$\+A_r \+s_r = \+b_r$} are the corresponding colluder equations for $r \in \{0,1\}$. If ${R_X \independent Y^{(1)} \cond \+ O, \+ X^{(1)} \setminus Y^{(1)}, \+ R \setminus R_X}$, then $p(R_X \cond \+ O, \+ X^{(1)}, (\+ R \setminus R_X) = 1)$ is identifiable if and only if ${\rank(\+A_r) = m}$ for $r = 0,1$.
\end{thm}
Proof. Assume first that ${\rank(\+A) = m}$. We first derive the identifying functional by writing
\[
  \begin{aligned}
    &p(R_X = r \cond \+ O, \+ X^{(1)}, (\+ R \setminus R_X) \eqs 1) \\
    &= p(R_X = r \cond \+ O, \+ X^{(1)} \setminus Y^{(1)}, (\+ R \setminus R_X) = 1) \\
    &= \frac{p(R_X \eqs r, X^{(1)} \cond \+ O, \+ X^{(1)} \setminus \{X^{(1)}, Y^{(1)}\}, (\+ R \setminus R_X) = 1)}{\sum_{r' = 0}^1 p(R_X \eqs r', X^{(1)} \cond \+ O, \+ X^{(1)} \setminus \{X^{(1)}, Y^{(1)}\}, (\+ R \setminus R_X) = 1)} \\
    &= \frac{p(R_X \eqs r, X^{(1)} \cond \+ O, \+ X^{(1)} \setminus \{X^{(1)}, Y^{(1)}\}, (\+ R \setminus \{R_X, R_Y\}) = 1, R_Y = 1)}{\sum_{r' = 0}^1 p(R_X \eqs r', X^{(1)} \cond \+ O, \+ X^{(1)} \setminus \{X^{(1)}, Y^{(1)}\}, (\+ R \setminus \{R_X, R_Y\}) = 1, R_Y = 1)} \\
    &= \left.\frac{p(R_X \eqs r, X^{(1)} \cond \+ Z, R_Y = 1)}{\sum_{r' = 0}^1 p(R_X \eqs r', X^{(1)} \cond \+ Z, R_Y = 1)}\right|_{\+ (\+ R \cap \+ Z) = 1},
  \end{aligned}
\]
where the first equality follows from the assumed conditional independence. The identifiability of the terms $p(R_X \eqs r, X^{(1)} \cond \+ Z, R_Y = 1)$ for $r \in \{0,1\}$ follows from the fact that a system of linear equations has a unique solution if $\+A_r$ is of full rank and this solution is given by the Moore--Penrose inverse $(\+A_r^\top \+A_r)^{-1} \+A_r^\top \+b_r$. Thus we obtain the identifying functional
\[
  p(R_X = r \cond \+ O, \+ X^{(1)}, (\+ R \setminus R_X) = 1)\rvert_{X^{(1)} = x_j} = \frac{s_{rj}}{s_{0j} + s_{1j}},
\]
where
\[
  \+s_r = \begin{pmatrix} s_{r1} \\ \vdots \\ s_{rm} \end{pmatrix} =
  \begin{pmatrix}
    p(R_X \eqs r, X^{(1)} \eqs x_1 \cond R_Y \eqs 1, \+ Z) \\
    \vdots \\
    p(R_X \eqs r, X^{(1)} \eqs x_m \cond R_Y \eqs 1, \+ Z)
  \end{pmatrix} = (\+A_r^\top \+A_r)^{-1} \+A_r^\top \+b_r.
\]
Assume then that $p(R_X \cond \+ O, \+ X^{(1)}, (\+ R \setminus R_X) = 1)$ is identifiable. If now ${\rank(\+A_r) < m}$, the system {$\+A_r \+s_r= \+b_r$} has infinitely many solutions. Therefore, there exist two models that differ in the target probabilities but are equal in all observed distributions. This is a contradiction and thus ${\rank(\+A_r) = m}$.

Note that if $m = q$ and ${\rank(\+A_r) = m}$ then $(\+A_r^\top \+A_r)^{-1} \+A_r^\top \+b = \+A_r^{-1} \+b_r$. If $m < q$ and ${\rank(\+A_r) = m}$ then $(\+A_r^\top \+A_r)^{-1} \+A_r^\top \+b_r$ gives the exact solution.

Theorem~\ref{thm:CCMidentifiability} directly generalizes to missing data models with unmeasured confounders represented by acyclic directed mixed graphs (missing data ADMGs), because the same colluder equations hold in such models and the conditions of the theorem invokes only a conditional independence relation and the rank conditions of the colluder matrices.

With Theorem~\ref{thm:CCMidentifiability} we may reconsider the examples from Section~\ref{sec:models}. First, we consider the example of Section~\ref{sec:example1} where both $X^{(1)}$ and $Y^{(1)}$ are binary in the missing data DAG of Figure~\ref{fig:colluder2}. In this DAG, the required conditional independence of Theorem~\ref{thm:CCMidentifiability} (and Lemma~\ref{lem:colluder_eqs_id}) reduces to $R_X \independent Y^{(1)} \cond X^{(1)}, R_Y$, which holds in the missing data DAG as $X^{(1)}$ d-separates the collider path that is open due to the conditioning on $R_Y$. Note that the solution provided by Theorem~\ref{thm:CCMidentifiability} is different from the one presented in Equation~\eqref{eq:binarymatrix}, which solves $p(R_Y = 1 \cond X^{(1)}, R_X = 0)$ instead of $p(R_X \cond X^{(1)}, Y^{(1)}, R_Y)$. Both strategies lead to identifying functionals in this instance. If instead $X^{(1)}$ is ternary and $Y^{(1)}$ is binary as in the example of Section~\ref{sec:example2}, the conditional independence condition is equally satisfied, but $m = 3$, $q = 2$, $\+A_r$ is a $2 \times 3$ matrix, and $\rank(\+A_r) < m$ implying that $p(R_X \cond X^{(1)},Y^{(1)},R_Y = 1)$ is not identifiable.

\subsection{Identification of the Full Law}

As a tool for full law identification with categorical variables, Theorem~\ref{thm:CCMidentifiability} should be applied to all colluder structures that are present in the missing data DAG when the number of categories of the true variables is known or assumed. If there are colluder structures where the conditional distribution of the colluding response indicator cannot be identified (either because the required conditional independence does not hold or the required rank conditions do not hold), then the full law cannot not identified via Theorem~\ref{thm:CCMidentifiability}. If all conditional distributions of colluding response indicators are identified and there are no self-censoring edges for non-colluding response indicators, then the full law is identified and the identifying functional can be derived. It turns out that the converse is also true, and the full law is not identifiable if Theorem~\ref{thm:CCMidentifiability} does not apply to all colluders in the missing data DAG. The following result characterizes these observations.

\begin{thm}\label{thm:CCMfulllaw}
A full law $p(\+R, \+X^{(1)}, \+ O)$ of a missing data DAG $\sG$ that is a CCM with respect to every colluder of $\sG$ is identifiable if and only if $\sG$ does not contain self-censoring edges and Theorem~\ref{thm:CCMidentifiability} applies for each colluder of $\sG$ such that the rank conditions hold. The identifying functional for the missingness mechanism $p(\+ R \cond \+ O, \+ X^{(1)})$ is given by the odds ratio parameterization of Equation~\eqref{eq:or_factorization} where the identifying functionals for the conditional distributions of colluding response indicators are provided by Theorem~\ref{thm:CCMidentifiability}.
\end{thm}
Proof. Assume first that there are no self-censoring edges and Theorem~\ref{thm:CCMidentifiability} applies such that the rank conditions hold for each colluder of $\sG$. The claim now follows as an immediate consequence of \citet[][Theorems~1 and 2]{pmlr-v119-nabi20a} as all terms on the right-hand side of Equation~\eqref{eq:or_factorization} are identifiable. Assume then that there exists a self-censoring edge. The non-identifiability of the full law follows again by \citet{pmlr-v119-nabi20a}. If there are no self-censoring edges, but there exists a colluder for which Theorem~\ref{thm:CCMidentifiability} applies but the required rank condition does not hold, then non-identifiability is implied by Theorem~\ref{thm:CCMidentifiability}. Suppose then that there exists a colluder for which the conditional independence assumption of Theorem~\ref{thm:CCMidentifiability} does not apply. Now it is possible to construct two models that agree on the observed law but disagree on the full law. We prove this using a parameter-counting argument in the Supplementary Material while also providing an explicit construction for $\text{CCM}(2,2)$.

We note that scenarios where the full law is a CCM with respect to only some but not all of the colluders of the missing data DAG fall under the purview of Lemma~\ref{lem:colluder_lemma}, rendering the full law non-identifiable \citep{pmlr-v115-bhattacharya20b}.

\section{Examples} \label{sec:examples}

\begin{figure}[t]
  \begin{center}
    \begin{subfigure}{0.32\textwidth}
      \begin{center}
        \begin{tikzpicture}[scale=1.5]
        \node [obs = {X}{X^{(1)}}] at (0,1) {};
        \node [obs = {Y}{Y^{(1)}}] at (1,1) {};
        \node [obs = {RX}{R_X}] at (0,0) {};
        \node [obs = {RY}{R_Y}] at (1,0) {};
        %\node [obs = {X*}{X}] at (0,-1) {};
        %\node [obs = {Y*}{Y}] at (1,-1) {};
        %
        \draw[->] (X) -- (Y);
        %\draw[->] (X) to[bend right=25] (X*);
        %\draw[->] (Y) to[bend left=25] (Y*);
        \draw[->] (X) -- (RY);
        \draw[->] (RX) -- (RY);
        %\draw[->] (RX) -- (X*);
        %\draw[->] (RY) -- (Y*);
        \draw[<->, dashed] (X) to[bend left=40] (Y);
        \draw[<->, dashed] (RX) to[bend right=40] (RY);
        \end{tikzpicture}
      \end{center}
      \caption{}
      \label{fig:example1}
    \end{subfigure}
    \begin{subfigure}{0.32\textwidth}
      \begin{center}
        \begin{tikzpicture}[scale=1.5]
        \node [obs = {X}{X^{(1)}}] at (0,1) {};
        \node [obs = {Y}{Y^{(1)}}] at (1,1) {};
        \node [obs = {RX}{R_X}] at (0,0) {};
        \node [obs = {RY}{R_Y}] at (1,0) {};
        %\node [obs = {X*}{X}] at (0,-1) {};
        %\node [obs = {Y*}{Y}] at (1,-1) {};
        %
        \draw[->] (X) -- (Y);
        %\draw[->] (X) to[bend right=25] (X*);
        %\draw[->] (Y) to[bend left=25] (Y*);
        \draw[->] (X) -- (RY);
        \draw[->] (RX) -- (RY);
        %\draw[->] (RX) -- (X*);
        %\draw[->] (RY) -- (Y*);
        \draw[<->, dashed] (RX) to[bend right=40] (RY);
        \draw[<->, dashed] (X) to[bend left=25] (RY);
        \end{tikzpicture}
        \end{center}
        \caption{}
        \label{fig:example2}
        \end{subfigure}
    \begin{subfigure}{0.32\textwidth}
      \begin{center}
        \begin{tikzpicture}[scale=1.5]
        \node [obs = {X}{X^{(1)}}] at (0,1) {};
        \node [obs = {Y}{Y^{(1)}}] at (1,1) {};
        \node [obs = {RX}{R_X}] at (0,0) {};
        \node [obs = {RY}{R_Y}] at (1,0) {};
        \draw[->] (X) -- (Y);
        \draw[->] (X) -- (RY);
        \draw[->] (RX) -- (RY);
        \draw[<->, dashed] (X) to[bend left=40] (Y);
        \draw[<->, dashed] (X) to[bend left=25] (RY);
        \end{tikzpicture}
      \end{center}
      \caption{}
      \label{fig:example3}
    \end{subfigure}

    \begin{subfigure}{0.32\textwidth}
      \begin{center}
        \begin{tikzpicture}[scale=1.5]
        \node [obs = {X}{X^{(1)}}] at (0,1) {};
        \node [obs = {Y}{Y^{(1)}}] at (1,1) {};
        \node [obs = {Z}{Z^{(1)}}] at (2,1) {};
        \node [obs = {RX}{R_X}] at (0,0) {};
        \node [obs = {RY}{R_Y}] at (1,0) {};
        \node [obs = {RZ}{R_Z}] at (2,0) {};
        \draw[->] (X) -- (Y);
        \draw[->] (Y) -- (Z);
        \draw[->] (X) -- (RY);
        \draw[->] (Y) -- (RZ);
        \draw[->] (RX) -- (RY);
        \draw[->] (RY) -- (RZ);
        \end{tikzpicture}
      \end{center}
      \caption{}
      \label{fig:example4}
    \end{subfigure}
    \begin{subfigure}{0.32\textwidth}
      \begin{center}
        \begin{tikzpicture}[scale=1.5]
        \node [obs = {X}{X^{(1)}}] at (0,1) {};
        \node [obs = {Y}{Y^{(1)}}] at (1,1) {};
        \node [obs = {Z}{Z^{(1)}}] at (2,1) {};
        \node [obs = {RX}{R_X}] at (0,0) {};
        \node [obs = {RY}{R_Y}] at (1,0) {};
        \node [obs = {RZ}{R_Z}] at (2,0) {};
        \draw[->] (X) -- (Y);
        \draw[->] (Z) -- (Y);
        \draw[->] (X) -- (RY);
        \draw[->] (Z) -- (RY);
        \draw[->] (RX) -- (RY);
        \draw[->] (RZ) -- (RY);
        \end{tikzpicture}
      \end{center}
      \caption{}
      \label{fig:example5}
   \end{subfigure}
   \begin{subfigure}{0.32\textwidth}
    \begin{center}
      \begin{tikzpicture}[scale=1.5]
      \node [obs = {X}{X^{(1)}}] at (0,1) {};
      \node [obs = {Y}{Y^{(1)}}] at (1,1) {};
      \node [obs = {Z}{Z^{(1)}}] at (-1,1) {};
      \node [obs = {RX}{R_X}] at (0,0) {};
      \node [obs = {RY}{R_Y}] at (1,0) {};
      \node [obs = {RZ}{R_Z}] at (-1,0) {};
      \draw[->] (X) -- (Y);
      \draw[->] (X) -- (Z);
      \draw[->] (X) -- (RY);
      \draw[->] (X) -- (RZ);
      \draw[->] (RX) -- (RY);
      \draw[->] (RX) -- (RZ);
      \end{tikzpicture}
    \end{center}
    \caption{}
    \label{fig:example6}
  \end{subfigure}
  \end{center}
  \caption{Examples of missing data ADMGs and DAGs with colluder structures. Theorem~\ref{thm:CCMidentifiability} can be applied in graphs (a), (b), and (d)--(f) but not in (c). Dashed bidirected edges denote the effects of unmeasured confounders. Observed proxy variables are omitted from the graphs for clarity.}
  \label{fig:examples}
\end{figure}
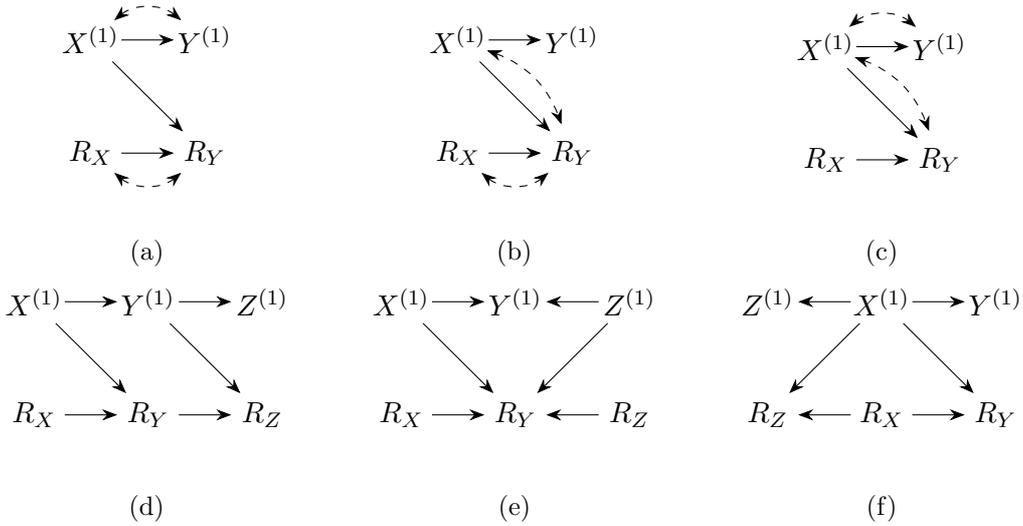

Next, we consider some additional examples including scenarios with unmeasured confounding. Figure~\ref{fig:examples} presents examples of missing data ADMGs and DAGs with colluder structures. Theorem~\ref{thm:CCMidentifiability} can be directly applied in the graphs Figures~\ref{fig:example1} and \ref{fig:example2} because the required conditional independence relations hold. In the graph of Figure~\ref{fig:example3} conditioning on $X^{(1)}$ and $R_Y$ opens the path $Y^{(1)} \leftrightarrow X^{(1)} \rightarrow R_Y \leftarrow R_X$, implying that Theorem~\ref{thm:CCMidentifiability} cannot be applied.

In the graph of Figure~\ref{fig:example4}, there is a sequential structure of colluders where $\{X^{(1)}, R_X\}$ is a colluder of $R_X$ and $\{Y^{(1)}, R_Y\}$ is a colluder of $R_Z$. The identifiability of the full law $p(X^{(1)},Y^{(1)},Z^{(1)},R_X,R_Y,R_Z)$ can be assessed via Theorem~\ref{thm:CCMfulllaw}, i.e., by applying Theorem~\ref{thm:CCMidentifiability} twice in this instance. First, we apply Theorem~\ref{thm:CCMidentifiability} to the first colluder ${X^{(1)} \rightarrow R_Y \leftarrow R_X}$. We have that $R_X \independent Y^{(1)} \cond X^{(1)}, Z^{(1)}, R_Y, R_Z$, thus $p(R_X|X^{(1)},Y^{(1)},Z^{(1)},R_Y = 1, R_Z = 1)$ is identifiable (assuming the rank conditions hold for the colluder matrices). Next, we apply Theorem~\ref{thm:CCMidentifiability} to the second colluder ${Y^{(1)} \rightarrow R_Z \leftarrow R_Y}$. We have that $R_Y \independent Z^{(1)} \cond X^{(1)}, Y^{(1)}, R_X, R_Z$, thus $p(R_Y|X^{(1)},Y^{(1)},Z^{(1)},R_X = 1, R_Z = 1)$ is also identifiable (assuming again the necessary rank conditions for the colluder matrices). There are no colluders of $R_X$ or self-censoring edges, making the full law is identifiable. Figures~\ref{fig:example5} and \ref{fig:example6} illustrate two analogous scenarios to Figure~\ref{fig:example4} with two simultaneous colluders of $R_Y$, and $\{R_X, X^{(1)}\}$ forming two colluders simultaneously, respectively. In both cases, the required conditional independence relation holds again, and the full law can be identified (assuming the necessary rank conditions).

\section{Simulation Study} \label{sec:simulation}

We study the estimation of the full law in categorical colluder models with simulated data. Estimation of the full law is a challenging task, especially in non-parametric or semi-parametric settings. Naive estimation via the solution to the colluder equations is possible for the conditional distributions of colluding response indicators, but there are no guarantees that the produced estimates are probabilities, or even positive with finite data. A more feasible approach is to directly employ a parameterization of the full law that relies on the OR parameterization of the missingness mechanism as defined in Equation~\ref{eq:or_factorization} \citep{Malinsky_2021}. We take a different approach and apply likelihood-based inference to the observed data law instead, thus avoiding the estimation of the OR terms in the factorization of Equation~\ref{eq:or_factorization}. We factorize the observed data law according to the structure of the missing data DAG $\sG$, analogously to the factorization of Equation~\eqref{eq:factorization}. Let $\vec{\+ v}^{\text{obs}} = (\+ v^{\text{obs}}_j)_{j=1}^n = (\+ o_j, \+ x_j, \+ r_j)_{j=1}^n$ be a data sample of $n$ observations. We denote the full law as
\[
  f(\+ v\, ; \bm\theta) = \prod_{V_i \in \+ V} \left.p(V_i \cond \pa[\sG](V_i) \,; \bm\theta)\right|_{\+ V = \+ v},
\]
where $\bm\theta$ is a vector of parameters that defines the full law and $p(\cdot \cond \cdot\,; \bm\theta)$ denotes a conditional probability parameterized by $\bm\theta$, and let
\[
  \begin{aligned}
  &s(\+ v^{\text{obs}}) = s(\+ o, \+ x, \+ r)= \\
  & \begin{cases}
    \{(\+ o, \+ x, \+ x, \+ r)\} & \text{if } x_k \neq \text{NA for all } k, \\
    \left\{ (\tilde{\+o}, \tilde{\+x}^{(1)}, \+ x, \tilde{\+r}) \in \val(\+ V) \middle| \tilde{\+o} \eqs \+ o, \tilde{\+r} \eqs \+ r, \tilde{x}^{(1)}_k \eqs x_k \text{ for all } x_k \neq \text{NA} \right\} & \text{otherwise},
  \end{cases}
  \end{aligned}
\]
where $\val(\+ V)$ denotes the state space of $\+ V$. The sum over the set $s(\+ v^{\text{obs}}$ represents integration over missing data. The likelihood function can now be written as
\begin{equation} \label{eq:likelihood}
  L(\bm\theta\,; \vec{\+ v}^{\text{obs}}) = \prod_{j = 1}^n p(\+ V^{\text{obs}} = \+ v^{\text{obs}}_j\,; \bm\theta) = \prod_{j=1}^n \sum_{\+ v \in s(\+ v^{\text{obs}}_j)} f(\+ v\, ; \bm\theta).
\end{equation}

We consider the missing data DAG of Figure~\ref{fig:colluder2} in our simulation study under two scenarios: one with binary $X^{(1)}$ and binary $Y^{(1)}$, and another where both $X^{(1)}$ and  $Y^{(1)}$ are quaternary. In both scenarios, the models are parametrized directly by the conditional probabilities. The parameters for $p(X^{(1)})$ and $p(Y^{(1)} \cond X^{(1)})$ are generated randomly while ensuring that $X^{(1)}$ and $Y^{(1)}$ have a moderate dependency. We fix  $p(R_X=1)=0.8$ and generate $p(R_Y=1 \cond X^{(1)}, R_X)$ randomly from the interval $[0.7, 0.9]$ so that the probabilities to observe $Y^{(1)}$ are different for different values of $X^{(1)}$ and $R_X$. The sample sizes are 1000, 10000, and 100000, and the number of simulation runs is 1000 for each sample size and scenario.

\begin{table}[!ht]
  \begin{center}
  \begin{tabular}{llcccc}
  \multicolumn{6}{l}{\bf Scenario $m=2$, $q=2$} \\
   & & \multicolumn{2}{c}{Mean bias} & \multicolumn{2}{c}{RMSE}\\
  Parameters for & $n$ & Min & Max & Mean & Max\\
  \hline
  $p(R_Y=1 \cond X^{(1)}, R_X)$                           & $1000$     &  $-0.0036$  & $\hphantom{-}0.0003$    &  $0.0702$  &  $0.1211$ \\
  $p(R_Y=1 \cond X^{(1)}, R_X)$                           & $10000$    &  $-0.0021$  & $\hphantom{-}0.0019$    &  $0.0255$  &  $0.0460$ \\
  $p(R_Y=1 \cond X^{(1)}, R_X)$                           & $100000$   &  $-0.0001$  & $\hphantom{-}0.0006$    &  $0.0076$  &  $0.0135$ \\
  $p(R_X=1)$, $p(X^{(1)})$, $p(Y^{(1)} \cond X^{(1)})$    & $1000$     &  $-0.0008$  & $-0.0001$   &  $0.0184$  &  $0.0233$ \\
  $p(R_X=1)$, $p(X^{(1)})$, $p(Y^{(1)} \cond X^{(1)})$    & $10000$    &  $-0.0002$  & $\hphantom{-}0.0004$  &  $0.0058$  &  $0.0072$ \\
  $p(R_X=1)$, $p(X^{(1)})$, $p(Y^{(1)} \cond X^{(1)})$    & $100000$   &  $-0.0000$  & $\hphantom{-}0.0000$  &  $0.0018$  &  $0.0022$
  \end{tabular}
  \\[0.6cm]
  \begin{tabular}{llcccc}
  \multicolumn{6}{l}{\bf Scenario $m=4$, $q=4$} \\
   & & \multicolumn{2}{c}{Mean bias} & \multicolumn{2}{c}{RMSE}\\
  Parameters for & $n$ & Min & Max & Mean & Max\\
  \hline
  $p(R_Y=1 \cond X^{(1)}, R_X)$                        & $1000$    &  $-0.0178$  &  $0.0045$  &  $0.1137$  &  $0.1944$ \\
  $p(R_Y=1 \cond X^{(1)}, R_X)$                        & $10000$   &  $-0.0035$  &  $0.0006$  &  $0.0548$  &  $0.1040$ \\
  $p(R_Y=1 \cond X^{(1)}, R_X)$                        & $100000$  &  $-0.0025$  &  $0.0015$  &  $0.0186$  &  $0.0350$ \\
  $p(R_X=1)$, $p(X^{(1)})$, $p(Y^{(1)} \cond X^{(1)})$ & $1000$    &  $-0.0017$  &  $0.0036$  &  $0.0388$  &  $0.0530$ \\
  $p(R_X=1)$, $p(X^{(1)})$, $p(Y^{(1)} \cond X^{(1)})$ & $10000$   &  $-0.0006$  &  $0.0003$  &  $0.0090$  &  $0.0145$ \\
  $p(R_X=1)$, $p(X^{(1)})$, $p(Y^{(1)} \cond X^{(1)})$ & $100000$  &  $-0.0001$  &  $0.0001$  &  $0.0028$  &  $0.0043$ 
  \end{tabular}
  \end{center}
  \caption{\label{tab:simulation_results}
  The results of the simulation study with scenarios $\textrm{CCM}(2,2)$  and $\textrm{CCM}(4,4)$. The average bias and RMSE are calculated for each parameter and their minimum, maximum and mean are reported for two groups of parameters.} 
\end{table}

We apply numerical maximum likelihood estimation and write the likelihood function of the full law according to Equation~\eqref{eq:likelihood}. The same likelihood could also be used in Bayesian estimation. We derive the gradient of the log-likelihood using symbolic derivation \citep{Deriv} and apply gradient-based numerical optimization to obtain the maximum likelihood estimates $\hat{\bm\theta}$. The simulation codes are available from GitHub at https://anonymous.4open.science/r/colluder-8DD1 (the repository is anonymized for reviewing).

The results of the simulations are presented in Table~\ref{tab:simulation_results}. The results are presented separately for the parameters of probabilities $p(R_Y=1 \cond X^{(1)}, R_X)$ and the other parameters because the parameters in the former group are expected to be more difficult to estimate than the parameters in the latter group due to the colluder structure. We calculated the average bias and the root mean square error (RMSE) for each parameter. Table~\ref{tab:simulation_results} presents the mean, the minimum, and the maximum of these statistics within the parameter groups.

All estimates appear to be unbiased. It can be seen from RMSEs that reliable estimation of the probabilities $p(R_Y=1 \cond X^{(1)}, R_X)$ requires large sample sizes. As expected, the estimation is more difficult in the quaternary scenario than the binary scenario.

\section{Application} \label{sec:application}

As a practical example, we estimate the full law from the data collected in Finnish Bachelor's Graduate Survey 2016 \citep{kandikysely}. The survey was sent to all students who had completed their Bachelor's degree to collect information on progress with studies, financing of studies, well-being, and balance between studies and other aspects of life. The dataset contains answers from 11708 graduates. We focus on two questions on the financing of studies. In the financing section, the graduates are first asked if they funded their studies by a student loan. The possible answers are 1 = Mainly, 2 = Partially, 3 = Not at all. $15\%$ of answers are missing. Later the graduates are asked if they funded their studies by personal income through work with same options for the answer. $8\%$ of answers are missing for this question.

\begin{table}[!ht]
\begin{center}
\begin{tabular}{ll}
Parameter & Estimate ($95\%$ CI) \\
\hline
$p(X^{(1)}=2)$                  & $0.446$ $(0.436, 0.456)$ \\
$p(X^{(1)}=3)$                  & $0.397$ $(0.387, 0.406)$ \\
$p(Y^{(1)}=2 | X^{(1)}=1)$      & $0.536$ $(0.510, 0.562)$ \\
$p(Y^{(1)}=3 | X^{(1)}=1)$      & $0.098$ $(0.082, 0.114)$ \\
$p(Y^{(1)}=2 | X^{(1)}=2)$      & $0.515$ $(0.500, 0.529)$ \\
$p(Y^{(1)}=3 | X^{(1)}=2)$      & $0.034$ $(0.029, 0.040)$ \\
$p(Y^{(1)}=2 | X^{(1)}=3)$      & $0.425$ $(0.410, 0.440)$ \\
$p(Y^{(1)}=3 | X^{(1)}=3)$      & $0.078$ $(0.070, 0.085)$ \\
$p(R_X=1)$                      & $0.853$ $(0.847, 0.860)$ \\
$p(R_Y=1 | X^{(1)}=1, R_X = 0)$ & $0.000$ \\
$p(R_Y=1 | X^{(1)}=2, R_X = 0)$ & $0.769$ \\
$p(R_Y=1 | X^{(1)}=3, R_X = 0)$ & $0.685$ \\
$p(R_Y=1 | X^{(1)}=1, R_X = 1)$ & $0.901$ $(0.887, 0.916)$ \\
$p(R_Y=1 | X^{(1)}=2, R_X = 1)$ & $0.967$ $(0.962, 0.973)$ \\
$p(R_Y=1 | X^{(1)}=3, R_X = 1)$ & $0.996$ $(0.994, 0.997)$ \\
\end{tabular}
\end{center}
\caption{\label{tab:bachelor}
Maximul likelihood estimates and their 95\% CIs for the categorical colluder model parameters of the Finnish Bachelor's Graduate Survey 2016. Value assignments $X^{(1)}=1$, $Y^{(1)}=1$, $R_X=0$, and $R_Y=0$ are the reference levels. The CIs are not reported for the probabilities $p(R_Y=1 \cond X^{(1)}=x, R_X = 0)$, $x \in \{1,2,3\}$ because they cannot be reliably estimated.}
\end{table}

We hypothesize that these questions could be modeled by a $\text{CCM}(3,3)$ where $X^{(1)}$ is the question on student loan and $Y^{(1)}$ is the question on income through work. It is natural to assume that $X^{(1)}$ and $Y^{(1)}$ are associated because they represent complementary forms of financing. The questions are not particularly sensitive, which gives some support for omitting self-censoring edges $X^{(1)} \rightarrow R_X$ and $Y^{(1)} \rightarrow R_Y$. As the question on student loan precedes the question on income through work, both $X^{(1)}$ and $R_X$ may affect $R_Y$. Especially, a graduate who has already reported that a student loan was the main form of financing may skip the further questions on financing.

Maximum likelihood estimation was applied using the approach described in Section~\ref{sec:simulation}. The estimates and their 95\% confidence intervals (CI) are summarized in Table~\ref{tab:bachelor}. It turns out that the data contains only little information on the probabilities $p(R_Y=1 \cond X^{(1)}=x, R_X = 0)$, $x \in \{1,2,3\}$. The probability $p(R_Y=1 \cond X^{(1)}=1, R_X = 0)$ is estimated to be very close to zero, which is unrealistic and may indicate that the modeling assumptions do not hold exactly.

\section{Conclusion} \label{sec:conclusion}

In this paper, we reassessed the role of the crucial colluder structure in terms of full law identifiability in missing data models represented by DAGs. We provided a necessary and sufficient graphical criterion for scenarios where the true variables related to the colluders are assumed to be categorical. This criterion also provides the identifying functional via the solution to the colluder equations. We also observed that typical constructions using binary variables to prove non-identifiability \citep{10.5555/1597348.1597382, 10.5555/1390681.1442797, NEURIPS2019_d88518ac} are not always adequate in missing data problems.

The two structures that prevent identifiability in missing data DAGs are self-censoring edges and colluders \citep{pmlr-v119-nabi20a}. \citet{li2023self} studied assumptions that enable the identification of the full law under self-censoring without colluders whereas we considered assumptions that permit the identification despite colluders but without self-censoring. An open question is whether there are assumptions that warrant identifying the full law when both self-censoring and colluders are present. We note that like the conditional independence assumption of Theorem~\ref{thm:CCMidentifiability}, assumptions regarding the presence of self-censoring edges or colluders are untestable.

\citet{pmlr-v119-nabi20a} also considered a more general form of the colluder structure, called a \emph{colluding path}, in the context of missing data ADMGs. They showed that the existence of a colluding path renders the full law non-parametrically non-identifiable in a missing data ADMG, analogously as colluders do in missing data DAGs. It is possible that the results presented in this paper for categorical variables related to colluders could also be extended to colluding paths in missing data ADMGs. We note however, that our approach to full law identification when colluders are present is applicable to missing data ADMGs as well if no colluding paths are present.

For some applications, identification of the full law may be excessive and the target law could suffice instead. While we focused on full law identifiability, which is a sufficient but not necessary condition for the identifiability of the target law, it is possible that categorical variables could play an important role in target law identification specifically, even when the full law is not identifiable.

The categorical colluder model exemplifies the difference between identification and estimation. Although the identification results are based on the OR parametrization and matrix equations, they are not explicitly needed in maximum likelihood estimation or Bayesian estimation providing that the likelihood is parametrized according the underlying graph. Theorems~\ref{thm:CCMidentifiability} and \ref{thm:CCMfulllaw} are still necessary for the estimation because they license the use of likelihood based inference.

% References
\bibliography{bibliography}

\newpage

\appendix

\setlength{\tabcolsep}{0.4em}

\setlength\tabcolsep{4pt}

\section{Details of the Identifiable Binary Colluder Model}

Below, we provide the general construction for the missing data DAG of Figure~1b of the main paper with binary variables under the assumption that $Y^{(1)}$ depends on $X^{(1)}$.

\begin{minipage}{0.19\linewidth}
\begin{tikzpicture}[scale=1.5]
\node [obs = {X}{X^{(1)}}] at (0,1) {};
\node [obs = {Y}{Y^{(1)}}] at (1,1) {};
\node [obs = {RX}{R_X}] at (0,0) {};
\node [obs = {RY}{R_Y}] at (1,0) {};
\node [obs = {X*}{X}] at (0,-1) {};
\node [obs = {Y*}{Y}] at (1,-1) {};
\draw[->] (X) -- (Y);
\draw[->, lightgray] (X) to[bend right=25] (X*);
\draw[->, lightgray] (Y) to[bend left=25] (Y*);
\draw[->] (X) -- (RY);
\draw[->] (RX) -- (RY);
\draw[->, lightgray] (RX) -- (X*);
\draw[->, lightgray] (RY) -- (Y*);
\end{tikzpicture}
\end{minipage}
\begin{minipage}{0.16\linewidth}
\scriptsize
\begin{tabular}{|c|c|}
\hline
$R_X$ & $p(R_X)$ \\
\hline
0     & $a$ \\
1     & $1 - a$ \\
\hline
\end{tabular}

\vspace{0.5cm}
\begin{tabular}{|c|c|}
\hline
$X^{(1)}$ & $p(X^{(1)})$ \\
\hline
0 & $b$ \\
1 & $1 - b$ \\
\hline
\end{tabular}
\end{minipage}
\begin{minipage}{0.27\linewidth}
\scriptsize
\begin{tabular}{|c|c|c|}
\hline
$Y^{(1)}$ & $X^{(1)}$ & $p(Y^{(1)}|X^{(1)})$ \\
\hline
0 & 0 & $c$ \\
1 & 0 & $1 - c$ \\
0 & 1 & $h$ \\
1 & 1 & $1 - h$ \\
\hline
\end{tabular}
\end{minipage}
\begin{minipage}{0.39\linewidth}
\scriptsize
\begin{tabular}{|c|c|c|c|}
\hline
$R_Y$ & $R_X$ & $X^{(1)}$ & $p(R_Y|R_X,X^{(1)})$ \\
\hline
0 & 0 & 0 & $d$ \\
1 & 0 & 0 & $1 - d$ \\
0 & 1 & 0 & $e$ \\
1 & 1 & 0 & $1 - e$ \\
0 & 0 & 1 & $f$ \\
1 & 0 & 1 & $1 - f$ \\
0 & 1 & 1 & $g$ \\
1 & 1 & 1 & $1 - g$ \\
\hline
\end{tabular}
\end{minipage}
\begin{center}
\scriptsize
\begin{tabular}{|c|c|c|c|c|c|c|c|}
\hline
$R_X$ & $R_Y$ & $X^{(1)}$ & $Y^{(1)}$ & $p(X^{(1)}, Y^{(1)}, R_X, R_Y)$ & $X$ & $Y$ & $p(X, Y, R_X, R_Y)$ \\
\hline
\multirow{4}{*}{0} & \multirow{4}{*}{0} & 0 & 0 & $abcd$                 & \multirow{4}{*}{\text{NA}} & \multirow{4}{*}{\text{NA}} & \multirow{4}{*}{$a\left[bd + (1-b)f\right]$} \\
                   &                    & 1 & 0 & $a(1-b)hf$             &                            &                            & \\
                   &                    & 0 & 1 & $ab(1-c)d$             &                            &                            & \\
                   &                    & 1 & 1 & $a(1-b)(1-h)f$         &                            &                            & \\
\hline\hline
\multirow{4}{*}{1} & \multirow{4}{*}{0} & 0 & 0 & $(1-a)bce$             & \multirow{2}{*}{0}         & \multirow{4}{*}{\text{NA}} & \multirow{2}{*}{$(1-a)be$} \\
                   &                    & 1 & 0 & $(1-a)(1-b)hg$         &                            &                            & \\
                   &                    & 0 & 1 & $(1-a)b(1-c)e$         & \multirow{2}{*}{1}         &                            & \multirow{2}{*}{$(1-a)(1-b)g$} \\
                   &                    & 1 & 1 & $(1-a)(1-b)(1-h)g$     &                            &                            & \\
\hline\hline
\multirow{4}{*}{0} & \multirow{4}{*}{1} & 0 & 0 & $abc(1-d)$             & \multirow{4}{*}{\text{NA}} & \multirow{2}{*}{0}         & \multirow{2}{*}{$a\left[bc(1-d) + (1-b)h(1-f)\right]$} \\
                   &                    & 1 & 0 & $a(1-b)h(1-f)$         &                            &                            & \\
                   &                    & 0 & 1 & $ab(1-c)(1-d)$         &                            & \multirow{2}{*}{1}         & \multirow{2}{*}{$a\left[b(1\!-\!c)(1\!-\!d) + (1\!-\!b)(1\!-\!h)(1\!-\!f)\right]$} \\
                   &                    & 1 & 1 & $a(1-b)(1-h)(1-f)$     &                            &                            & \\
\hline\hline
\multirow{4}{*}{1} & \multirow{4}{*}{1} & 0 & 0 & $(1-a)bc(1-e)$         & 0                          & 0                          & $(1-a)bc(1-e)$ \\
                   &                    & 1 & 0 & $(1-a)(1-b)h(1-g)$     & 1                          & 0                          & $(1-a)(1-b)h(1-g)$ \\
                   &                    & 0 & 1 & $(1-a)b(1-c)(1-e)$     & 0                          & 1                          & $(1-a)b(1-c)(1-e)$ \\
                   &                    & 1 & 1 & $(1-a)(1-b)(1-h)(1-g)$ & 1                          & 1                          & $(1-a)(1-b)(1-h)(1-g)$ \\
\hline
\end{tabular}
\end{center}
Let $r = p(Y = 0, R_X = 0, R_Y = 1)$ and $s = p(Y = 1, R_X = 0, R_Y = 1)$. From positivity it follows $a,b,c,d,e,f,g,h,r,s \in (0,1)$, and dependency of $X^{(1)}$ and $Y^{(1)}$ implies that $c \neq h$.

We have the following system of equations
\[
  \begin{aligned}
  a\left[bc(1-d) + (1-b)h(1-f)\right] &= r,\\
  a\left[b(1\!-\!c)(1\!-\!d) + (1\!-\!b)(1\!-\!h)(1\!-\!f)\right] &= s.
  \end{aligned}
\]
Solving these equations in terms of $d$ and $f$ provides us with the following formulas
\[
  \begin{aligned}
    1-d &= p(R_Y = 1 \cond X^{(1)} = 0, R_X = 0) = \frac{r - h r - h s}{ab(c - h)}, \\
    1-f &= p(R_Y = 1 \cond X^{(1)} = 1, R_X = 0) = \frac{r - c r - c s}{a(b-1)(c-h)}.
  \end{aligned}
\]
It remains to show that the remaining parameters are identifiable from the observed data distribution. Clearly, $a$ is identifiable directly from the observed data distribution. The parameter $b$ is identified by noting that $X^{(1)} \independent R_X$ and we obtain
\[
  b = p(X^{(1)} = 0) = p(X^{(1)} \cond R_X = 1) = p(X \cond R_X = 1).
\]
Similarly, parameters $c$ and $h$ are identified by noting that $Y^{(1)} \independent \{R_X,R_Y\} \cond X^{(1)}$ so we can write
\[
  \begin{aligned}
    c &= p(Y^{(1)} = 0 \cond X^{(1)} = 0) \\
      &= p(Y^{(1)} = 0 \cond X^{(1)} = 0, R_X = 1, R_Y = 1) \\
      &= p(Y = 0 \cond X = 0, R_X = 1, R_Y = 1), \\
    h &= p(Y^{(1)} = 0 \cond X^{(1)} = 1) \\
      &= p(Y^{(1)} = 0 \cond X^{(1)} = 1, R_X = 1, R_Y = 1) \\
      &= p(Y = 0 \cond X = 1, R_X = 1, R_Y = 1).
  \end{aligned}
\]
Finally, $e$ and $g$ are identified as
\[
  \begin{aligned}
    e &= \frac{p(X = 0, R_X = 1, R_Y = 0)}{(1-a)b}, \\
    g &= \frac{p(X = 1, R_X = 1, R_Y = 0)}{(1-a)(1-b)}.
  \end{aligned}
\]
Thus $d$ and $f$ are also identifiable making the full law identifiable.

\newpage

\section{Details of the Non-Identifiable Ternary Colluder Model}

We show that when $X^{(1)}$ is ternary and $Y^{(1)}$ is binary, it is possible to construct two missing data models that agree on the observed data law but disagree on the full law in the missing data DAG of Figure~1b of the main paper with binary variables under the dependency of $X^{(1)}$ and $Y^{(1)}$. The general construction is shown below. We denote $m = 1 - b - c$ for short.

\vspace{0.25cm}
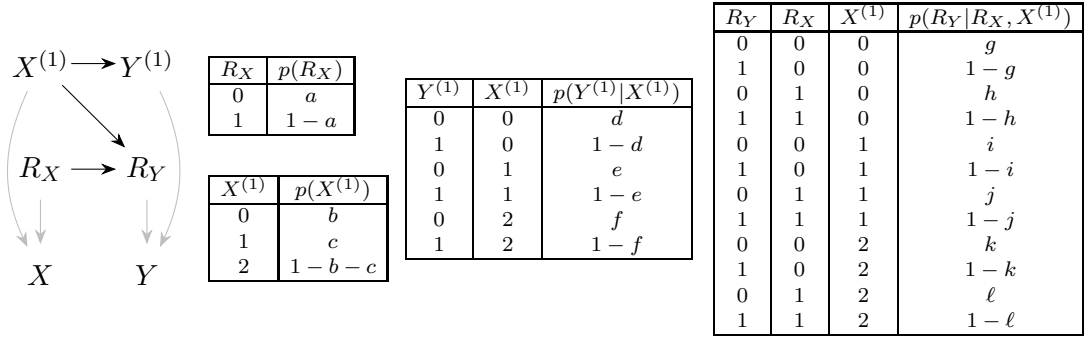
\begin{minipage}{0.18\linewidth}
\begin{tikzpicture}[scale=1.4]
\node [obs = {X}{X^{(1)}}] at (0,1) {};
\node [obs = {Y}{Y^{(1)}}] at (1,1) {};
\node [obs = {RX}{R_X}] at (0,0) {};
\node [obs = {RY}{R_Y}] at (1,0) {};
\node [obs = {X*}{X}] at (0,-1) {};
\node [obs = {Y*}{Y}] at (1,-1) {};
\draw[->] (X) -- (Y);
\draw[->,lightgray] (X) to[bend right=25] (X*);
\draw[->,lightgray] (Y) to[bend left=25] (Y*);
\draw[->] (X) -- (RY);
\draw[->] (RX) -- (RY);
\draw[->, lightgray] (RX) -- (X*);
\draw[->, lightgray] (RY) -- (Y*);
\end{tikzpicture}
\end{minipage}
\begin{minipage}{0.17\linewidth}
\scriptsize
\begin{tabular}{|c|c|}
\hline
$R_X$ & $p(R_X)$ \\
\hline
0     & $a$ \\
1     & $1 - a$ \\
\hline
\end{tabular}

\vspace{0.5cm}
\begin{tabular}{|c|c|}
\hline
$X^{(1)}$ & $p(X^{(1)})$ \\
\hline
0  & $b$ \\
1  & $c$ \\
2  & $1 - b - c$ \\
\hline
\end{tabular}
\end{minipage}
\begin{minipage}{0.27\linewidth}
\scriptsize
\begin{tabular}{|c|c|c|}
\hline
$Y^{(1)}$ & $X^{(1)}$ & $p(Y^{(1)}|X^{(1)})$ \\
\hline
0 & 0 & $d$ \\
1 & 0 & $1 - d$ \\
0 & 1 & $e$ \\
1 & 1 & $1 - e$ \\
0 & 2 & $f$ \\
1 & 2 & $1 - f$ \\
\hline
\end{tabular}
\end{minipage}
\begin{minipage}{0.39\linewidth}
\scriptsize
\begin{tabular}{|c|c|c|c|}
\hline
$R_Y$ & $R_X$ & $X^{(1)}$ & $p(R_Y|R_X,X^{(1)})$ \\
\hline
0 & 0 & 0 & $g$ \\
1 & 0 & 0 & $1 - g$\\
0 & 1 & 0 & $h$ \\
1 & 1 & 0 & $1 - h$\\
0 & 0 & 1 & $i$ \\
1 & 0 & 1 & $1 - i$ \\
0 & 1 & 1 & $j$ \\
1 & 1 & 1 & $1 - j$\\
0 & 0 & 2 & $k$\\
1 & 0 & 2 & $1 - k$\\
0 & 1 & 2 & $\ell$\\
1 & 1 & 2 & $1 - \ell$\\
\hline
\end{tabular}
\end{minipage}
\begin{center}
\scriptsize
\begin{tabular}{|c|c|c|c|c|c|c|c|}
\hline
$R_X$ & $R_Y$ & $X^{(1)}$ & $Y^{(1)}$ & $p(X^{(1)}, Y^{(1)}, R_X, R_Y)$ & $X$ & $Y$ & $p(X, Y, R_X, R_Y)$ \\
\hline
\multirow{6}{*}{0} & \multirow{6}{*}{0} & 0 & 0 & $abdg$                & \multirow{6}{*}{\text{NA}} & \multirow{6}{*}{\text{NA}} & \multirow{6}{*}{$a\left[bg + ci + mk\right]$} \\
                   &                    & 1 & 0 & $acei$                &                            &                            & \\
                   &                    & 2 & 0 & $amfk$                &                            &                            & \\
                   &                    & 0 & 1 & $ab(1-d)g$            &                            &                            & \\
                   &                    & 1 & 1 & $ac(1-e)i$            &                            &                            & \\
                   &                    & 2 & 1 & $am(1-f)k$            &                            &                            & \\
\hline\hline
\multirow{6}{*}{1} & \multirow{6}{*}{0} & 0 & 0 & $(1-a)bdh$            & \multirow{2}{*}{0}         & \multirow{6}{*}{\text{NA}} & \multirow{2}{*}{$(1-a)bh$} \\
                   &                    & 1 & 0 & $(1-a)cej$            &                            &                            & \\
                   &                    & 2 & 0 & $(1-a)mf\ell$         & \multirow{2}{*}{1}         &                            & \multirow{2}{*}{$(1-a)cj$} \\
                   &                    & 0 & 1 & $(1-a)b(1-d)h$        &                            &                            & \\
                   &                    & 1 & 1 & $(1-a)c(1-e)j$        & \multirow{2}{*}{2}         &                            & \multirow{2}{*}{$(1-a)m\ell$} \\
                   &                    & 2 & 1 & $(1-a)m(1-f)\ell$     &                            &                            & \\
\hline\hline
\multirow{6}{*}{0} & \multirow{6}{*}{1} & 0 & 0 & $abd(1-g)$            & \multirow{6}{*}{\text{NA}} & \multirow{3}{*}{0}         & \multirow{3}{*}{$a\left[bd(1-g) + ce(1-i) + mf(1-k)\right]$} \\
                   &                    & 1 & 0 & $ace(1-i)$            &                            &                            & \\
                   &                    & 2 & 0 & $amf(1-k)$            &                            &                            & \\
                   &                    & 0 & 1 & $ab(1-d)(1-g)$        &                            & \multirow{3}{*}{1}         & \multirow{3}{*}{$a\left[b(1\!-\!d)(1\!-\!g) + c(1\!-\!e)(1\!-\!i) + m(1\!-\!f)(1\!-\!k)\right]$} \\
                   &                    & 1 & 1 & $ac(1-e)(1-i)$        &                            &                            & \\
                   &                    & 2 & 1 & $am(1-f)(1-k)$        &                            &                            & \\
\hline\hline
\multirow{6}{*}{1} & \multirow{6}{*}{1} & 0 & 0 & $(1-a)bd(1-h)$        & 0                          & 0                          & $(1-a)bd(1-h)$ \\
                   &                    & 1 & 0 & $(1-a)ce(1-i)$        & 1                          & 0                          & $(1-a)ce(1-i)$ \\
                   &                    & 2 & 0 & $(1-a)mf(1-\ell)$     & 2                          & 0                          & $(1-a)mf(1-\ell)$ \\
                   &                    & 0 & 1 & $(1-a)b(1-d)(1-h)$    & 0                          & 1                          & $(1-a)b(1-d)(1-h)$ \\
                   &                    & 1 & 1 & $(1-a)c(1-e)(1-i)$    & 1                          & 1                          & $(1-a)c(1-e)(1-i)$ \\
                   &                    & 2 & 1 & $(1-a)m(1-f)(1-\ell)$ & 2                          & 1                          & $(1-a)m(1-f)(1-\ell)$ \\
\hline
\end{tabular}
\end{center}
Let us now choose the values of the parameters $a,\ldots,\ell$ for the two models. Let the subscript denote the model, e.g., $a_1$ is the value of $a$ in model 1. Let $b_1 = b_2 = 0.25$ (i.e., $c_1 = c_2 = 0.5$), $d_1 = d_2 = f_1 = f_2 = n$, and $g_1 = g$, $g_2 = k$, $k_1 = k$, $k_2 = g$ for two values of $g$ and $k$ such that $g \neq k$. We let the values of the other parameters agree between the models and drop the subscript for such parameters. We confirm that the observed data laws agree
\[
  \begin{aligned}
    a_1&[b_1g_1 + c_1i_1 + m_1k_1] \\
    &= a[0.25g + ci + 0.25k] \\
    &= a[0.25k + ci + 0.25g] \\
    &= a_2[b_2g_2 + c_2i_2 + m_2k_2], \\[0.25cm]
    (1-a_1)b_1h_1    &= (1-a)\cdot 0.25h = (1-a_2)b_2h_2, \\
    (1-a_1)c_1j_1    &= (1-a)cj = (1-a_2)c_2j_2, \\
    (1-a_1)m_1\ell_1 &= (1-a)\cdot 0.25\ell = (1-a_2)m_2\ell_2, \\[0.25cm]
    a_1&[b_1d_1(1-g_1) + c_1e_1(1-i_1) + m_1f_1(1-k_1)] \\
    &= a[0.25n(1-g) + 0.5e(1-i) + 0.25n(1-k)] \\
    &= a[0.25n(1-k) + 0.5e(1-i) + 0.25n(1-g)] \\
    &= a[0.25d_2(1-k) + 0.5e(1-i) + 0.25f_2(1-g)] \\
    &= a_2[b_2d_2(1-g_2) + c_2e_2(1-i_2) + m_2f_2(1-k_2)],\\[0.25cm]
    a_1&[b_1(1-d_1)(1-g_1) + c(1-e_1)(1-i_1) + m_1(1-f_1)(1-k_1)] \\
    &= a[0.25(1-n)(1-g) + 0.5(1-e)(1-i) + 0.25(1-n)(1-k)] \\
    &= a[0.25(1-n)(1-k) + 0.5(1-e)(1-i) + 0.25(1-n)(1-g)] \\
    &= a[0.25(1-d_2)(1-k) + 0.5(1-e)(1-i) + 0.25(1-f_2)(1-g)] \\
    &= a[b_2(1-d_2)(1-g_2) + c_2(1-e_2)(1-i_2) + m_2(1-f_2)(1-k_2)], \\[0.25cm]
    (1-a_1)b_1d_1(1-h_1)        &= (1-a)\cdot 0.25n(1-h)       = (1-a_2)b_2d_2(1-h_2), \\
    (1-a_1)c_1e_1(1-i_1)        &= (1-a)\cdot 0.5e(1-i)        = (1-a_2)c_2e_2(1-i_2), \\
    (1-a_1)m_1f_1(1-\ell_1)     &= (1-a)\cdot 0.25n(1-\ell)    = (1-a_2)m_2f_2(1-\ell_2), \\
    (1-a_1)b_1(1-d_1)(1-h_1)    &= (1-a)\cdot 0.25(1-n)(1-h)   = (1-a_2)b_2(1-d_2)(1-h_2), \\
    (1-a_1)c_1(1-e_1)(1-i_1)    &= (1-a)\cdot 0.5(1-e)(1-i)    = (1-a_2)c_2(1-e_2)(1-i_2), \\
    (1-a_1)m_1(1-f_1)(1-\ell_1) &= (1-a)\cdot 0.5(1-n)(1-\ell) = (1-a_2)m_2(1-f_2)(1-\ell_2).
  \end{aligned}
\]
However, the full laws do not agree, as we can observe from the following probabilities
\[
  \begin{aligned}
    a_1b_1d_1g_1 = a\cdot0.25ng &\neq a\cdot0.25nk = a_2b_2d_2g_2, \\
    a_1m_1f_1k_1 = a\cdot0.25nk &\neq a\cdot0.25ng = a_2m_2f_2k_2, \\
    a_1b_1(1-d_1)g_1 = a\cdot0.25(1-n)g &\neq a\cdot0.25(1-n)k = a_2b_2(1-d_2)g_2, \\
    a_1m_1(1-f_1)k_1 = a\cdot0.25(1-n)k &\neq a\cdot0.25(1-n)g = a_2m_2(1-f_2)k_2, \\
    a_1b_1d_1(1-g_1) = a\cdot0.25n(1 - g) &\neq a\cdot0.25n(1 - k) = a_2b_2d_2(1-g_2), \\
    a_1m_1f_1(1-k_1) = a\cdot0.5n(1 - k) &\neq a\cdot0.5n(1 - g) = a_2m_2f_2(1-k_2), \\
    a_1b_1(1-d_1)(1-g_1) = a\cdot0.25(1-n)(1-g) &\neq a\cdot0.25(1-n)(1-k) = a_2b_2(1-d_2)(1-g_2), \\
    a_1m_1(1-f_1)(1-k_1) = a\cdot0.5(1-n)(1-k) &\neq a\cdot0.5(1-n)(1-g) = a_2m_2(1-f_2)(1-k_2).
  \end{aligned}
\]
Thus the full law is not identifiable.

\newpage

\section{Non-Identifiable Cross-Censoring Model (Theorem~2)}

We consider Theorem~2 of the main paper and the case where ${R_X \not\independent Y^{(1)} \cond \+ O, \+ X^{(1)} \setminus Y^{(1)}, \+ R \setminus R_X}$ for a colluder $\{R_X, X^{(1)}\}$ of $R_Y$. When this conditional independence does not hold, it must be the case that there is either a self-censoring edge $Y^{(1)} \rightarrow R_Y$ or a directed edge $Y^{(1)} \rightarrow R_X$, because conditioning on $\+ O, \+ X^{(1)} \setminus Y^{(1)}, \+ R \setminus R_X$ blocks all backdoor paths from $R_X$ to $Y^{(1)}$ and all collider paths of the form $R_X \rightarrow \cdots \rightarrow R_Z \leftarrow \cdots \leftarrow Y^{(1)}$ except the case of a self-censoring edge. Because there are no edges from members of $\+ R$ to vertices not in $\+ R$, there are no other possible paths. It remains to show that if the edge $Y^{(1)} \rightarrow R_X$ exists, then the full law is not identifiable. First, we consider a special case with binary variables, followed by a general argument.

\vspace{0.5cm}
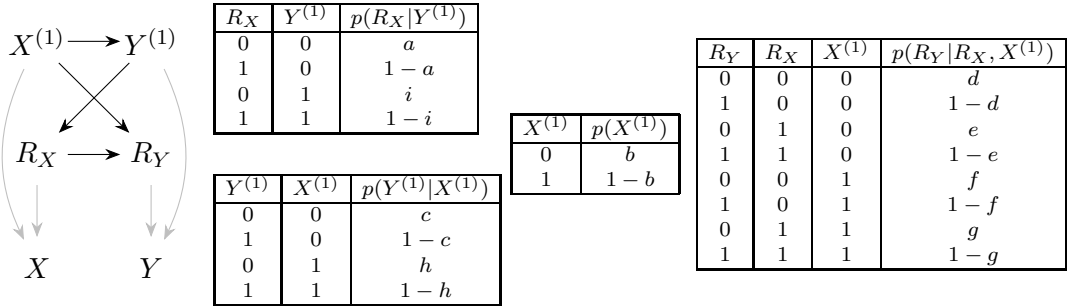
\begin{minipage}{0.19\linewidth}
  \begin{tikzpicture}[scale=1.5]
  \node [obs = {X}{X^{(1)}}] at (0,1) {};
  \node [obs = {Y}{Y^{(1)}}] at (1,1) {};
  \node [obs = {RX}{R_X}] at (0,0) {};
  \node [obs = {RY}{R_Y}] at (1,0) {};
  \node [obs = {X*}{X}] at (0,-1) {};
  \node [obs = {Y*}{Y}] at (1,-1) {};
  \draw[->] (X) -- (Y);
  \draw[->, lightgray] (X) to[bend right=25] (X*);
  \draw[->, lightgray] (Y) to[bend left=25] (Y*);
  \draw[->] (X) -- (RY);
  \draw[->] (Y) -- (RX);
  \draw[->] (RX) -- (RY);
  \draw[->, lightgray] (RX) -- (X*);
  \draw[->, lightgray] (RY) -- (Y*);
  \end{tikzpicture}
  \end{minipage}
  \begin{minipage}{0.26\linewidth}
  \scriptsize
  \begin{tabular}{|c|c|c|}
  \hline
  $R_X$ & $Y^{(1)}$ & $p(R_X|Y^{(1)})$ \\
  \hline
  0     & 0 & $a$ \\
  1     & 0 & $1 - a$ \\
  0     & 1 & $i$ \\
  1     & 1 & $1 - i$ \\
  \hline
  \end{tabular}
  
  \vspace{0.5cm}
  \begin{tabular}{|c|c|c|}
    \hline
    $Y^{(1)}$ & $X^{(1)}$ & $p(Y^{(1)}|X^{(1)})$ \\
    \hline
    0 & 0 & $c$ \\
    1 & 0 & $1 - c$ \\
    0 & 1 & $h$ \\
    1 & 1 & $1 - h$ \\
    \hline
  \end{tabular}
  \end{minipage}
  \begin{minipage}{0.16\linewidth}
  \scriptsize
  \begin{tabular}{|c|c|}
    \hline
    $X^{(1)}$ & $p(X^{(1)})$ \\
    \hline
    0 & $b$ \\
    1 & $1 - b$ \\
    \hline
    \end{tabular}
  \end{minipage}
  \begin{minipage}{0.38\linewidth}
  \scriptsize
  \begin{tabular}{|c|c|c|c|}
  \hline
  $R_Y$ & $R_X$ & $X^{(1)}$ & $p(R_Y|R_X,X^{(1)})$ \\
  \hline
  0 & 0 & 0 & $d$ \\
  1 & 0 & 0 & $1 - d$ \\
  0 & 1 & 0 & $e$ \\
  1 & 1 & 0 & $1 - e$ \\
  0 & 0 & 1 & $f$ \\
  1 & 0 & 1 & $1 - f$ \\
  0 & 1 & 1 & $g$ \\
  1 & 1 & 1 & $1 - g$ \\
  \hline
  \end{tabular}
\end{minipage}

\begin{center}
  \scriptsize
  \begin{tabular}{|c|c|c|c|c|c|c|c|}
  \hline
  $R_X$ & $R_Y$ & $X^{(1)}$ & $Y^{(1)}$ & $p(X^{(1)}, Y^{(1)}, R_X, R_Y)$ & $X$ & $Y$ & $p(X, Y, R_X, R_Y)$ \\
  \hline
  \multirow{4}{*}{0} & \multirow{4}{*}{0} & 0 & 0 & $abcd$                 & \multirow{4}{*}{\text{NA}} & \multirow{4}{*}{\text{NA}} & \multirow{4}{*}{$\begin{aligned}&a\left[bcd + (1-b)hf\right] \\ &+ i\left[b(1-c)d + (1-b)(1-h)f\right] \end{aligned}$} \\
                     &                    & 1 & 0 & $a(1-b)hf$             &                            &                            & \\
                     &                    & 0 & 1 & $ib(1-c)d$             &                            &                            & \\
                     &                    & 1 & 1 & $i(1-b)(1-h)f$         &                            &                            & \\
  \hline\hline
  \multirow{4}{*}{1} & \multirow{4}{*}{0} & 0 & 0 & $(1-a)bce$             & \multirow{2}{*}{0}         & \multirow{4}{*}{\text{NA}} & \multirow{2}{*}{$be\left[(1-a)c + (1-i)(1-c)\right]$} \\
                     &                    & 1 & 0 & $(1-a)(1-b)hg$         &                            &                            & \\
                     &                    & 0 & 1 & $(1-i)b(1-c)e$         & \multirow{2}{*}{1}         &                            & \multirow{2}{*}{$(1-b)g\left[(1-a)h + (1-i)(1-h)\right]$} \\
                     &                    & 1 & 1 & $(1-i)(1-b)(1-h)g$     &                            &                            & \\
  \hline\hline
  \multirow{4}{*}{0} & \multirow{4}{*}{1} & 0 & 0 & $abc(1-d)$             & \multirow{4}{*}{\text{NA}} & \multirow{2}{*}{0}         & \multirow{2}{*}{$a\left[bc(1-d) + (1-b)h(1-f)\right]$} \\
                     &                    & 1 & 0 & $a(1-b)h(1-f)$         &                            &                            & \\
                     &                    & 0 & 1 & $ib(1-c)(1-d)$         &                            & \multirow{2}{*}{1}         & \multirow{2}{*}{$i\left[b(1\!-\!c)(1\!-\!d) + (1\!-\!b)(1\!-\!h)(1\!-\!f)\right]$} \\
                     &                    & 1 & 1 & $i(1-b)(1-h)(1-f)$     &                            &                            & \\
  \hline\hline
  \multirow{4}{*}{1} & \multirow{4}{*}{1} & 0 & 0 & $(1-a)bc(1-e)$         & 0                          & 0                          & $(1-a)bc(1-e)$ \\
                     &                    & 1 & 0 & $(1-a)(1-b)h(1-g)$     & 1                          & 0                          & $(1-a)(1-b)h(1-g)$ \\
                     &                    & 0 & 1 & $(1-i)b(1-c)(1-e)$     & 0                          & 1                          & $(1-i)b(1-c)(1-e)$ \\
                     &                    & 1 & 1 & $(1-i)(1-b)(1-h)(1-g)$ & 1                          & 1                          & $(1-i)(1-b)(1-h)(1-g)$ \\
  \hline
  \end{tabular}
\end{center}

Next, we choose the values of the parameters $a,\ldots,i$ for the two models. Let the subscript denote the model, e.g., $a_1$ is the value of $a$ in model 1. For simplicity, we choose specific numeric values for all parameters, but note that there are infinitely many parametrizations that agree of the observed data law but disagree on the full law. The chosen values are
\[
  \begin{aligned}
    a_1 &= \frac{229}{400}      &  a_2 &= \frac{283}{492} \\
    b_1 &= \frac{59}{114}       &  b_2 &= \frac{108}{209} \\
    c_1 &= \frac{40}{59}        &  c_2 &= \frac{41}{60} \\
    d_1 &= \frac{1028}{1145}    &  d_2 &= \frac{1074}{1415} \\
    e_1 &= \frac{1}{3}          &  e_2 &= \frac{1}{3} \\
    f_1 &= \frac{7373}{11450}   &  f_2 &= \frac{10949}{14150} \\
    g_1 &= \frac{2}{5}          &  g_2 &= \frac{2}{5} \\
    h_1 &= \frac{80}{99}        &  h_2 &= \frac{82}{101} \\
    i_1 &= \frac{1}{10}         &  i_2 &= \frac{1}{12}
  \end{aligned}
\]
These choices produce the following probabilities for the observed data law in both models
\begin{center}
  \begin{tabular}{|c|c|c|}
  \hline
  $X$ & $Y$ & $p(X, Y, R_X, R_Y)$ \\
  \hline
  NA & NA   & $69/200$ \\
  0  & NA   & $1/10$ \\
  1  & NA   & $1/10$ \\
  NA & 0    & $1/10$ \\
  NA & 1    & $1/200$ \\
  0  & 0    & $1/10$ \\
  1  & 0    & $1/10$ \\
  0  & 1    & $1/10$ \\
  1  & 1    & $1/20$ \\
  \hline
  \end{tabular}
\end{center}
but the full laws disagree, for example:
\[
  a_1b_1c_1d_1 = \frac{257}{1425} \neq \frac{1611}{10450} = a_2b_2c_2d_2,
\]
thus the full law is not identifiable.

Now suppose that the missing data model is a $\text{CCM}(m, q)$ with respect to the colluder. Let us first consider the case that the edge $X^{(1)} \rightarrow Y^{(1)}$ exists. The full law has $mq + 2m + q - 1$ parameters whearas an upper bound for the parameters of the observed law is $mq + m + q$ (i.e., the number of cells in the observed law column of the table above minus one). The difference is $m - 1$, thus there are infinitely many mappings from the full law to the observed law.

Finally, we consider the case where there edge $X^{(1)} \rightarrow Y^{(1)}$ does not exist. Now, the number of parameters of the full law is $3m + 2q - 2$. The naive upper bound for the number of parameters of the observed law is no longer sufficient. Instead, we use the following parametrization:
\[
  \begin{aligned}
  & &p(R_X = 0, Y = y_i | R_Y = 1) & & &i = 1,\ldots,q - 1 \\
  & &p(Y = y_i | R_Y = 1)          & & &i = 1,\ldots,q - 1 \\
  & &p(R_X = 0) & & & \\
  & &p(R_Y = 0, X = x_j | R_X = 1) & & &j = 1,\ldots,m - 1 \\
  & &p(X = x_j | R_X = 1)            & & &j = 1,\ldots, m - 1 \\
  & &p(R_Y = 0 | R_X = 0) & & & \\
  & &p(R_Y = 0 | R_X = 1) & & &
  \end{aligned}
\]
leading to $2m + 2q - 1$ parameters, with a difference of $m - 1$ parameters, thus there are again infinitely many mappings. The observed law is recovered from the parameters as follows:
\[
  \begin{aligned}
  p&(X = \text{NA}, Y = \text{NA}, R_X = 0, R_Y = 0) \\
    &= p(R_X = 0, R_Y = 0) \\
    &= p(R_Y = 0 | R_X = 0)p(R_X = 0) \\[0.25cm]
  p&(X = x_j, Y = \text{NA}, R_X = 1, R_Y = 0) \\
    &= p(X = x_j, R_X = 1, R_Y = 0) \\
    &=  p(X = x_j, R_Y = 0 | R_X = 1)p(R_X = 1) \\
    &=  p(X = x_j, R_Y = 0 | R_X = 1)(1 - p(R_X = 0)) \\[0.25cm]
  p&(Y = y_i, X = \text{NA}, R_X = 0, R_Y = 1) \\
    &= p(Y = y_i, R_X = 0, R_Y = 1) \\
    &=  p(Y = y_i, R_X = 0 | R_Y = 1)p(R_Y = 1) \\[0.25cm]
    &=  p(Y = y_i, R_X = 0 | R_Y = 1) \sum_{r=0}^1 (1 - p(R_Y = 0|R_X = r))p(R_X = r) \\[0.25cm]
  p&(X = x_j, Y = y_i, R_X = 1, R_Y = 1) \\
    &= p(Y = y_i | X = x_j, R_X = 1 R_Y = 1)p(X = x_j, R_X = 1, R_Y = 1) \\
    &= p(Y = y_i | R_X = 1, R_Y = 1)p(X = x_j, R_X = 1, R_Y = 1) \\
    &= p(Y = y_i | R_X = 1, R_Y = 1)p(X = x_j, R_Y = 1|R_X = 1)p(R_X = 1) \\
    &= \frac{p(Y = y_i, R_X = 1 | R_Y = 1)}{p(R_X = 1 | R_Y = 1)}p(X = x_j, R_Y = 1|R_X = 1)p(R_X = 1) \\
  \end{aligned}
\]
for $i = 1\ldots,q$ and $j = 1,\ldots,m-1$. The case with $j = m$ is obtained by replacing $p(X = x_j, R_Y = 0 | R_X = 1)$ with $p(R_Y = 0 | R_X = 1) - \sum_{j=1}^{m-1} p(X = x_j, R_Y = 0 | R_X = 1)$ and similarly the case with $i = q$ by replacing $p(Y = y_i, R_X = 0 | R_Y = 1)$ with $p(R_X = 0 | R_Y = 1) - \sum_{i = 1}^{q-1} p(Y = y_i, R_X = 0 | R_Y = 1)$. Similarly $p(Y = y_i, R_X = 1 | R_Y = 1)$ is obtained by writing $p(Y = y_i|R_Y = 1) - p(Y = y_i, R_X = 0 | R_Y = 1)$ and $p(X = x_j, R_Y = 1|R_X = r)$ is obtained by writing $p(R_Y = 0 | R_X = r) - p(X = x_j, R_Y = 0|R_X = r)$. Finally, we used the conditional independence $Y^{(1)} \independent X^{(1)} \mid R_X,R_Y$ to obtain $p(Y = y_i | R_X = 1, R_Y = 1)$ from $p(Y = y_i| X = x_j, R_X = 1, R_Y = 1)$ (note that $Y = Y^{(1)}$ and $X = X^{(1)}$ when conditioning on $R_Y = 1$ and $R_X = 1$).

\end{document}